\documentclass{article}%
\usepackage{amsfonts}
\usepackage{amsmath}
\usepackage{amssymb}
\usepackage{graphicx}%
\providecommand{\U}[1]{\protect\rule{.1in}{.1in}}
\newtheorem{theorem}{Theorem}
\newtheorem{acknowledgement}[theorem]{Acknowledgement}

\newtheorem{definition}[theorem]{Definition}

\newenvironment{proof}[1][Proof]{\noindent\textbf{#1.} }{\ \rule{0.5em}{0.5em}}

\begin{document}

\title{Einstein-Hilbert action with cosmological term from Chern-Simons gravity}
\author{N. Gonz\'{a}lez, P. Salgado, G. Rubio, \ S. Salgado\\Departamento de F\'{\i}sica, Universidad de Concepci\'{o}n \\Casilla 160-C, Concepci\'{o}n, Chile}
\maketitle

\begin{abstract}
We propose a modification to the Lie algebra $S$-expansion method. The
modification is carried out by imposing a condition on the $S$-expansion
procedure, when the semigroup is given by a cyclic group of even order. The
$S$-expanded algebras are called $S_{H}$-expanded algebras where $S=Z_{2n}$.
\ The invariant tensors for $S_{H}$-expanded algebras are calculated and
the\ dual formulation of $S_{H}$-expansion procedure is proposed. We consider
the $S_{H}$-expansion of the five-dimensional $AdS$ algebra and its
corresponding invariants tensors are found. Then a Chern-Simons Lagrangian
invariant under the five-dimensional $AdS$ algebra $S_{H}$-expanded is
constructed and its relationship to the general relativity is studied. \footnote[2]{\textcopyright 2016. This manuscript version is made available under the CC-BY-NC-ND 4.0 license http://creativecommons.org/licenses/by-nc-nd/4.0/}

\end{abstract}

\section{\textbf{Introduction}}

The Lie algebra expansion procedure was introduced for the first time in
Ref.~\cite{hat}, and subsequently studied in general in Refs.~\cite{azcarr1},
\cite{azcarr2}, \cite{azcarr3}. \ The expansion procedure is different from
the In\"{o}n\"{u}-Wigner contraction method \cite{paperiw} albeit, when the
algebra dimension does not change in the process, it may lead to a simple
In\"{o}n\"{u}-Wigner or generalized In\"{o}n\"{u}-Wigner contraction in the
sense of Weimar-Woods \cite{paperww1}, \cite{paperww2}, \cite{paperww3}.
Furthermore, the algebras to which it leads have in general higher dimension
than the original one, in which case they cannot be related to it by any
contraction or deformation process. \ \ \ \ \ \ 

The expansion method proposed in Refs.\ \cite{hat}, \cite{azcarr1} consists in
considering the original algebra as described by its associated Maurer- Cartan
forms on the group manifold. Some of the group parameters are rescaled by a
factor $\lambda$, and the Maurer-Cartan forms are expanded as a power series
in $\lambda$. This series is finally truncated in a way that assures the
closure of the expanded algebra.

In Refs. \cite{exp1}, \cite{exp2}, \cite{exp3} was proposed a natural
outgrowth of power series expansion method, which is based on combining the
structure constant of the algebra $\mathfrak{g}$ with the inner law of a
semigroup $S$ in order to define the Lie bracket of a new $S$-expanded algebra.

Theorem 1 of Ref. \cite{azcarr1} shows that, in the more general case, the
expanded Lie algebra has the structure constants%
\[
C_{(A,i)(Bj)}^{\text{ \ \ \ \ \ \ \ \ \ }(C,k)}=\left\{
\begin{array}
[c]{c}%
0\text{ \ \ when \ }i+j\neq k\\
C_{AB}^{\text{ \ \ \ \ }C}\text{ \ when \ }i+j=k
\end{array}
\right.
\]
where $i,j,k=0,\cdot\cdot\cdot,N$ \ correspond to the order of the expansion,
and $N$ is the truncation order. These structure constants can also be
obtained within the $S$-expansion procedure. In order to achieve this, one
must consider the $0_{S}$-reduction of an $S$-expanded algebra where $S$
corresponds to the semigroup. The Maurer-Cartan forms power series expansion
of an algebra $\mathfrak{g}$, with truncation order $N$, coincides with the
$0_{S}$-reduction of the $S_{E}^{(N)}$-expanded algebra (see Ref.
\cite{exp1}). \ \ This is of course no coincidence. The set of powers of the
rescaling parameter $\lambda$, together with the truncation at order $N$,
satisfy precisely the multiplication law of $S_{E}^{(N)}$. As a matter of
fact, we have $\lambda^{\alpha}\lambda^{\beta}=\lambda^{\alpha+\beta}$ and the
truncation can be imposed as $\lambda^{\alpha}=0$ when $\alpha>N$. \ It is for
this reason that one must demand $0_{S}T_{A}=0$ in order to obtain the
Maurer-Cartan expansion as an $S_{E}^{(N)}$-expansion: in this case the zero
of the semigroup is the zero of the field as well.

The $S$-expansion procedure is valid no matter what the structure of the
original $\mathfrak{g}$ Lie algebra is, and in this sense it is very general.
However, when something about the structure of $\mathfrak{g}$ is known, a lot
more can be done. As an example, in the context of Maurer-Cartan expansion,
the rescaling and truncation can be performed in several ways depending on the
structure of $\mathfrak{g}$, leading to several kinds of expanded algebras.
Important examples of this are the generalized In\"{o}n\"{u}--Wigner
contraction, or the $M$ algebra as an expansion of $osp(32|1)$ (see Refs.
\cite{azcarr1}, \cite{azcarr2}). This is also the case in the context of
$S$-expansions. When some information about the structure of $\mathfrak{g}$ is
available, it is possible to find subalgebras of $\mathfrak{G}=S_{E}%
^{(N)}\times\mathfrak{g}$ and other kinds of reduced algebras. In this way,
all the algebras obtained by the Maurer-Cartan expansion procedure can be
reobtained. New kinds of $S$-expanded algebras can also be obtained by
considering semigroups different from $S_{E}^{(N)}$.

The purpose of this paper is to introduce a modification to the $S$-expansion
method. The modification is carried out by imposing a condition on the
$S$-expansion procedure, when the semigroup is given by a cyclic group of even
order. The $S$-expanded algebras are called $S_{H}$-expanded algebras where
$S=Z_{2n}$. \ The invariant tensors for $S_{H}$-expanded algebras are
calculated and the\ dual formulation of $S_{H}$-expansion procedure is proposed.

Following Ref.\cite{standardg}, we consider the $S_{H}$-expansion of the
five-dimensional $AdS$ algebra and its corresponding invariants tensors are
obtained. We also consider the study of a Chern-Simons Lagrangian invariant
under the five-dimensional $AdS$ algebra $S_{H}$-expanded and its relationship
to the general relativity.

This paper is organized as follows: In Section 2 we review some aspects of
the Lie algebra $S$-expansion procedure. In Section 3 it is introduced a
modification to the $S$-expansion method. In Section 4 the invariant
tensors for $S_{H}$-expanded algebras are obtained and the\ dual formulation
of $S_{H}$-expansion procedure is constructed. In Section 5 we consider the
$S_{H}$-expansion of the five-dimensional $AdS$ algebra and its corresponding
invariants tensors are found. In Section 6 a Chern-Simons Lagrangian
invariant under the five-dimensional $AdS$ algebra $S_{H}$-expanded is
constructed and its relationship to the general relativity is studied. A
comment about possible developments and an appendix concludes the work.\ 

\section{\textbf{The S-expansion procedure}}

In this section we shall review the main aspects of the $S$-expansion
procedure and their properties introduced in Ref. \cite{exp1}.

Let $S=\left\{  \lambda_{\alpha}\right\}  $ be an abelian semigroup with
2-selector $K_{\alpha\beta}^{\ \ \ \gamma}$ defined by
\begin{equation}
K_{\alpha\beta}^{\ \ \ \gamma}=\left\{
\begin{array}
[c]{cc}%
1 & \ \ \lambda_{\alpha}\lambda_{\beta}=\lambda_{\gamma}\\
0 & otherwise,
\end{array}
\right.
\end{equation}
and $\mathfrak{g}$ a Lie (super)algebra with basis $\left\{  T_{A}\right\}  $
and structure constant $C_{AB}^{\ \ \ C}$,
\begin{equation}
\left[  T_{A},T_{B}\right]  =C_{AB}^{\ \ \ C}T_{C}.
\end{equation}
Then it may be shown that the product $\mathfrak{G}=S\times\mathfrak{g}$ is
also a Lie (super)algebra with structure constants $C_{(A,\alpha)(B,\beta
)}^{\ \ \ \ \ \ \ \ \ \ \ \ (C,\gamma)}=K_{\alpha\beta}^{\ \ \gamma}%
C_{AB}^{\ \ \ \ C}$,
\begin{equation}
\left[  T_{(A,\alpha)},T_{(B,\beta)}\right]  =C_{(A,\alpha)(B,\beta
)}^{\ \ \ \ \ \ \ \ \ \ \ \ (C,\gamma)}T_{(C,\gamma)}.
\end{equation}
The proof is direct and may be found in Ref. \cite{exp1}.

\begin{definition}
Let $S$ be an abelian semigroup and $\mathfrak{g}$ a Lie algebra. The Lie
algebra $\mathfrak{G}$ defined by $\mathfrak{G}=S\times\mathfrak{g}$ is called
$S$-Expanded algebra of $\mathfrak{g}$.
\end{definition}

When the semigroup has a zero element $0_{S}\in S$, it plays a somewhat
peculiar role in the $S$-expanded algebra. The above considerations motivate
the following definition:

\begin{definition}
Let $S$ be an abelian semigroup with a zero element $0_{S}\in S$, and let
$\mathfrak{G}=S\times\mathfrak{g}$ be an $S$-expanded algebra. The algebra
obtained by imposing the condition $0_{S}\mathbf{T}_{A}=0$ on $\mathfrak{G}$
(or a subalgebra of it) is called $0_{S}$-reduced algebra of $\mathfrak{G}$
(or of the subalgebra).
\end{definition}

An $S$-expanded algebra has a fairly simple structure. Interestingly, there
are at least two ways of extracting smaller algebras from $S\times
\mathfrak{g}$. The first one gives rise to a \textit{resonant subalgebra},
while the second produces reduced algebras. \ In particular, a resonant
subalgebra can be obtained as follow.

Let $\mathfrak{g}=%
%TCIMACRO{\tbigoplus _{p\in I}}%
%BeginExpansion
{\textstyle\bigoplus_{p\in I}}
%EndExpansion
V_{p}$ be a decomposition of $\mathfrak{g}$ in subspaces $V_{p}$, where $I$ is
a set of indices. \ For each $p,q\in I$ it is always possible to define
$i_{\left(  p,q\right)  }\subset I$ such that%
\begin{equation}
\left[  V_{p},V_{q}\right]  \subset%
%TCIMACRO{\tbigoplus \limits_{r\in i_{\left(  p,q\right)  }}}%
%BeginExpansion
{\textstyle\bigoplus\limits_{r\in i_{\left(  p,q\right)  }}}
%EndExpansion
V_{r}. \label{eq33}%
\end{equation}
\textbf{ }Now, let $S=%
%TCIMACRO{\tbigcup _{p\in I}}%
%BeginExpansion
{\textstyle\bigcup_{p\in I}}
%EndExpansion
S_{p}$ be a subset decomposition of the abelian semigroup $S$ such that%
\begin{equation}
S_{p}\cdot S_{q}\subset%
%TCIMACRO{\tbigcup _{r\in i_{\left(  p,q\right)  }}}%
%BeginExpansion
{\textstyle\bigcup_{r\in i_{\left(  p,q\right)  }}}
%EndExpansion
S_{p}. \label{eq34}%
\end{equation}
When such subset decomposition $S=%
%TCIMACRO{\tbigcup _{p\in I}}%
%BeginExpansion
{\textstyle\bigcup_{p\in I}}
%EndExpansion
S_{p}$ exists, then we say that this decomposition is in resonance with the
subspace decomposition of $\mathfrak{g},$ $\mathfrak{g}=%
%TCIMACRO{\tbigoplus _{p\in I}}%
%BeginExpansion
{\textstyle\bigoplus_{p\in I}}
%EndExpansion
V_{p}$.

The resonant subset decomposition is crucial in order to systematically
extract subalgebras from the $S$-expanded algebra $\mathfrak{G}=S\times
\mathfrak{g}$, as is proven in the following\medskip

\textbf{Theorem IV.2 of Ref}. \cite{exp1}: Let $\mathfrak{g}=%
%TCIMACRO{\tbigoplus _{p\in I}}%
%BeginExpansion
{\textstyle\bigoplus_{p\in I}}
%EndExpansion
V_{p}$ be a subspace decomposition of $\mathfrak{g}$, with a structure
described by eq. $\left(  \ref{eq33}\right)  ,$ and let $S=%
%TCIMACRO{\tbigcup _{p\in I}}%
%BeginExpansion
{\textstyle\bigcup_{p\in I}}
%EndExpansion
S_{p}$ be a resonant subset decomposition of the abelian semigroup $S$, with
the structure given in eq. $\left(  \ref{eq34}\right)  $. Define the subspaces
of $\mathfrak{G}=S\times\mathfrak{g}$,%
\begin{equation}
W_{p}=S_{p}\times V_{p},\text{ \ }p\in I.
\end{equation}
Then,%
\begin{equation}
\mathfrak{G}_{R}=%
%TCIMACRO{\tbigoplus _{p\in I}}%
%BeginExpansion
{\textstyle\bigoplus_{p\in I}}
%EndExpansion
W_{p}%
\end{equation}
is a subalgebra of $\mathfrak{G}=S\times\mathfrak{g}$.

\textbf{Proof:} \ the proof may be found in Ref. \cite{exp1}.

\begin{definition}
The algebra $G_{R}=%
%TCIMACRO{\tbigoplus _{p\in I}}%
%BeginExpansion
{\textstyle\bigoplus_{p\in I}}
%EndExpansion
W_{p}$ obtained is called a Resonant Subalgebra of the $S$-expanded algebra
$\mathfrak{G}=S\times\mathfrak{g}$.
\end{definition}

A useful property of the $S$-expansion procedure is that it provides us with
an invariant tensor for the $S$-expanded algebra $\mathfrak{G}=S\times
\mathfrak{g}$ in terms of an invariant tensor for $\mathfrak{g}$. As shown in
Ref. \cite{exp1} the theorem VII.2 provide a general expression for an
invariant tensor for a $0_{S}$-reduced algebra.

\textbf{Theorem VII.2 of Ref. \cite{exp1}:} \ Let $S$ be an abelian semigroup
with nonzero elements $\lambda_{i}$, $i=0,\cdots,N$ and $\lambda_{N+1}=0_{S}$.
Let $\mathfrak{g}$ be a Lie (super)algebra of basis $\left\{  T_{A}\right\}
$, and let $\langle T_{A_{n}}\cdots T_{A_{n}}\rangle$ be an invariant tensor
for $\mathfrak{g}$. The expression
\begin{equation}
\langle T_{(A_{1},i_{1})}\cdots T_{(A_{n},i_{n})}\rangle=\alpha_{j}%
K_{i_{a}\cdots i_{n}}^{\ \ \ \ \ j}\langle T_{A_{1}}\cdots T_{A_{n}}\rangle
\end{equation}

where $\alpha_{j}$ are arbitrary constants, corresponds to an invariant tensor
for the $0_{S}$-reduced algebra obtained from $\mathfrak{G}=S\times
\mathfrak{g}$.

\textbf{Proof:} \ the proof may be found in section $4.5$ of Ref. \cite{exp1}.

\section{\textbf{ The }$S_{H}$\textbf{-expansion Method}}

In Ref. \cite{exp1} it was found that given a Lie algebra $\mathfrak{g}$ and a
semigroup $S,$ we can find a new Lie algebra $S\times\mathfrak{g}$ called
$S$-expanded Lie algebra . If the semigroup $S$ is endowed with a zero element
$0_{S}$, then it is possible to obtain a new Lie algebra by imposing the
condition $0_{S}\otimes T_{A}=0$ on the $S$-expanded algebra.

In this section we consider the expansion of a Lie algebra $\mathfrak{g}$ in
the case that the semigroup $S$ is given by a cyclic group with an even number
of elements $Z_{2n}=\left\{  \lambda_{0},\ldots,\lambda_{2n-1}\right\}  $. We
will prove that if we impose the condition $\lambda_{i}\otimes T_{A}%
+\lambda_{i+n}\otimes T_{A}=0$ on $Z_{2n}\times\mathfrak{g}$, then the
resulting structure is also a Lie algebra. The condition will be called
$H$-condition and the new Lie algebra obtained will be denoted by $\left(
Z_{2n}\times\mathfrak{g}\right)  _{H}$.

The motivation for imposing the $H$-condition is based on the fact that, when
the mechanism of $S$-expansion \cite{exp1} is used, then it can be shown that
the \ \ \ \ \ $Z_{2}$-expanded $\mathfrak{so}(3)$-algebra corresponds to an
algebra isomorphic to the $\mathfrak{so}(4)$ algebra
\begin{equation}
\mathfrak{so}(4)\simeq Z_{2}\times\mathfrak{so}(3). \label{eq:so3exp}%
\end{equation}
An open question is: is it possible to obtain, by S-expansion, the
$\mathfrak{so}(3,1)$ algebra from the $\mathfrak{so}(3)$ algebra?.

To try to answer this question, consider the $\mathfrak{so}(3,1)$ algebra
written in the form
\begin{equation}
\left[  J_{i},J_{j}\right]  =\varepsilon_{ij}^{\text{ \ }k}J_{k},\quad\left[
J_{i},K_{j}\right]  =\varepsilon_{ij}^{\text{ \ }k}K_{k},\quad\left[
K_{i},K_{j}\right]  =-\varepsilon_{ij}^{\text{ \ }k}J_{k}. \label{1}%
\end{equation}
The fundamental difference that makes the $\mathfrak{so}(4)$ and
$\mathfrak{so}(3,1)$ algebras are not isomorphic is the minus sign appears on
the right side of the commutator between the generators of the Lorentz boosts
$K_{i}$.

Consider the following set of generators $\left\{  J_{i},K_{i},A_{i}%
,B_{i}\right\}  $ where $A_{i}=-J_{i}$ and $B_{i}=-K_{i}$ and find the
commutation relations between them:
\begin{equation}%
\begin{array}
[c]{ll}%
\left[  J_{i},J_{j}\right]  =\varepsilon_{ij}{}^{k}J_{k},\quad & \left[
J_{i},K_{j}\right]  =\varepsilon_{ij}{}^{k}K_{k},\\
\left[  J_{i},A_{j}\right]  =\varepsilon_{ij}{}^{k}A_{k},\quad & \left[
J_{i},B_{j}\right]  =\varepsilon_{ij}{}^{k}B_{k},\\
\left[  K_{i},K_{j}\right]  =\varepsilon_{ij}{}^{k}A_{k},\quad & \left[
K_{i},A_{j}\right]  =\varepsilon_{ij}{}^{k}B_{k},\\
\left[  K_{i},B_{j}\right]  =\varepsilon_{ij}{}^{k}J_{k},\quad & \left[
A_{i},A_{j}\right]  =\varepsilon_{ij}{}^{k}J_{k},\\
\left[  A_{i},B_{j}\right]  =\varepsilon_{ij}{}^{k}K_{k},\quad & \left[
B_{i},B_{j}\right]  =\varepsilon_{ij}{}^{k}A_{k}.
\end{array}
\label{2}%
\end{equation}
These commutation relations coincide with the commutation relations of the
algebra $Z_{4}\times\mathfrak{so}(3)$. In fact, using the $S$-expansion method
with $S=Z_{4}=\left\{  \lambda_{0},\lambda_{1},\lambda_{2},\lambda
_{3}\right\}  $ and $J_{(i,\alpha)}=\lambda_{\alpha}J_{i}$ we have
\begin{equation}%
\begin{array}
[c]{ll}%
\left[  J_{(i,0)},J_{(j,0)}\right]  =\varepsilon_{ij}^{\text{ \ }k}%
J_{(k,0)},\quad & \left[  J_{(i,0)},J_{(j,1)}\right]  =\varepsilon
_{ij}^{\text{ \ }k}J_{(k,1)},\\
\left[  J_{(i,0)},J_{(j,2)}\right]  =\varepsilon_{ij}^{\text{ \ }k}%
J_{(k,2)},\quad & \left[  J_{(i,0)},J_{(j,3)}\right]  =\varepsilon
_{ij}^{\text{ \ }k}J_{(k,3)},\\
\left[  J_{(i,1)},J_{(j,1)}\right]  =\varepsilon_{ij}^{\text{ \ }k}%
J_{(k,2)},\quad & \left[  J_{(i,1)},J_{(j,2)}\right]  =\varepsilon
_{ij}^{\text{ \ }k}J_{(k,3)},\\
\left[  J_{(i,1)},J_{(j,3)}\right]  =\varepsilon_{ij}^{\text{ \ }k}%
J_{(k,0)},\quad & \left[  J_{(i,2)},J_{(j,2)}\right]  =\varepsilon
_{ij}^{\text{ }k}J_{(k,0)},\\
\left[  J_{(i,2)},J_{(j,3)}\right]  =\varepsilon_{ij}^{\text{ }k}%
J_{(k,1)},\quad & \left[  J_{(i,3)},J_{(j,3)}\right]  =\varepsilon
_{ij}^{\text{ \ }k}J_{(k,2)},
\end{array}
\label{3}%
\end{equation}
the commutation relations (\ref{2}) and (\ref{3}) coincide under the
correspondence $J_{(i,0)}=J_{i}$, $J_{(i,1)}=K_{i}$, $J_{(i,2)}=A_{i}$,
$J_{(i,3)}=B_{i}$. The above result shows that if we apply the conditions
\begin{equation}
J_{(i,2)}=-J_{(i,0)},\quad J_{(i,3)}=-J_{(i,1)} \label{4}%
\end{equation}
on the expanded algebra $Z_{4}\times\mathfrak{so}(3)$ we obtain the algebra
$\mathfrak{so}(3,1)$.

It is interesting that these conditions are not independent. The second
condition can be obtained if we multiply the first by $\lambda_{1}$, according
to the law of multiplication of $Z_{4}$. Operating with any element $Z_{4}$
get one any of the two conditions given in (\ref{4}).

An interesting question is: do we get a Lie algebra if we apply this condition
on arbitrary Lie algebra $Z_{4}\times\mathfrak{g}$ ?. To answer, we consider
the $Z_{4}$-expansion of a Lie algebra $\mathfrak{g}$ . Using the S-expansion
method with $S=Z_{4}=\left\{  \lambda_{0},\lambda_{1},\lambda_{2},\lambda
_{3}\right\}  $ and $T_{(A,\alpha)}=\lambda_{\alpha}T_{A}$ we have
\begin{equation}%
\begin{array}
[c]{ll}%
\left[  T_{(A,0)},T_{(B,0)}\right]  =C_{AB}{}^{C}T_{(C,0)},\quad & \left[
T_{(A,0)},T_{(B,1)}\right]  =C_{AB}{}^{C}T_{(C,1)},\\
\left[  T_{(A,0)},T_{(B,2)}\right]  =C_{AB}{}^{C}T_{(C,2)},\quad & \left[
T_{(A,0)},T_{(B,3)}\right]  =C_{AB}{}^{C}T_{(C,3)},\\
\left[  T_{(A,1)},T_{(B,1)}\right]  =C_{AB}{}^{C}T_{(C,2)},\quad & \left[
T_{(A,1)},T_{(B,2)}\right]  =C_{AB}{}^{C}T_{(C,3)},\\
\left[  T_{(A,1)},T_{(B,3)}\right]  =C_{AB}{}^{C}T_{(C,0)},\quad & \left[
T_{(A,2)},T_{(B,2)}\right]  =C_{AB}{}^{C}T_{(C,0)},\\
\left[  T_{(A,2)},T_{(B,3)}\right]  =C_{AB}{}^{C}T_{(C,1)},\quad & \left[
T_{(A,3)},T_{(B,3)}\right]  =C_{AB}{}^{C}T_{(C,2)}.
\end{array}
\label{5}%
\end{equation}
Applying the conditions
\begin{equation}
T_{(A,2)}=-T_{(A,0)},\quad T_{(A,3)}=-T_{(A,1)} \label{cond1}%
\end{equation}
over (\ref{5}), the following commutation relations are obtained
\begin{equation}%
\begin{array}
[c]{ll}%
\left[  T_{(A,0)},T_{(B,0)}\right]  =C_{AB}{}^{C}T_{(C,0)},\quad & \left[
T_{(A,0)},T_{(B,1)}\right]  =C_{AB}{}^{C}T_{(C,1)},\\
\left[  T_{(A,0)},\left(  -T_{(B,0)}\right)  \right]  =C_{AB}{}^{C}\left(
-T_{(C,0)}\right)  ,\quad & \left[  T_{(A,0)},\left(  -T_{(B,1)}\right)
\right]  =C_{AB}{}^{C}\left(  -T_{(C,1)}\right)  ,\\
\left[  T_{(A,1)},T_{(B,1)}\right]  =C_{AB}{}^{C}\left(  -T_{(C,0)}\right)
,\quad & \left[  T_{(A,1)},\left(  -T_{(B,0)}\right)  \right]  =\left(
-T_{(C,1)}\right)  ,\\
\left[  T_{(A,1)},\left(  -T_{(B,1)}\right)  \right]  =C_{AB}{}^{C}%
T_{(C,0)},\quad & \left[  \left(  -T_{(A,0)}\right)  ,\left(  -T_{(B,0)}%
\right)  \right]  =C_{AB}{}^{C}T_{(C,0)},\\
\left[  \left(  -T_{(A,0)}\right)  ,\left(  -T_{(B,1)}\right)  \right]
=C_{AB}{}^{C}T_{(C,1)},\quad & \left[  \left(  -T_{(A,1)}\right)  ,\left(
-T_{(B,1)}\right)  \right]  =`C_{A}B^{C}`\left(  -T_{(C,0)}\right)  ,
\end{array}
\label{6}%
\end{equation}
which can be rewritten as
\begin{align}
\left[  T_{(A,0)},T_{(B,0)}\right]   &  =C_{AB}{}^{C}T_{(C,0)},\quad\left[
T_{(A,0)},T_{(B,1)}\right]  =C_{AB}{}^{C}T_{(C,1)},\nonumber\\
\quad\left[  T_{(A,1)},T_{(B,1)}\right]   &  =-C_{AB}{}^{C}T_{(C,0)}.
\label{eq:z4xg}%
\end{align}

To verify that this algebra corresponds to a Lie algebra, we define $T_{\mu
}=\left(  T_{(A,0)},T_{(A,1)}\right)  $, with $\mu=1,\ldots,2m$ and
$m=\dim\mathfrak{g,}$ whereupon the commutation relations (\ref{eq:z4xg})
become $\left[  T_{\mu},T_{\nu}\right]  =f_{\mu\nu}{}^{\rho}T_{\rho}$, with
\begin{align}
f_{AB}{}^{C}  &  =C_{AB}{}^{C},\nonumber\\
f_{AB}{}^{C+m}  &  =0,\nonumber\\
f_{A(B+m)}{}^{C}=f_{(A+m)B}{}^{C}  &  =0,\nonumber\\
f_{A(B+m)}{}^{C+m}=f_{(A+m)B}^{\text{ \ \ \ \ \ \ \ \ \ }C+m}  &  =C_{AB}%
{}^{C},\label{7}\\
f_{(A+m)(B+m)}{}^{C}  &  =-C_{AB}{}^{C},\nonumber\\
f_{(A+m)(B+m)}{}^{C+m}  &  =0.\nonumber
\end{align}
From (\ref{7}) one can see that the structure constants are antisymmetric in
their low indices. It is straightforward to verify that they satisfy the
Jacobi identity. In effect,
\begin{align}
&  f_{\mu\nu}{}^{\lambda}f_{\lambda\rho}{}^{\varepsilon}+f_{\nu\rho}%
{}^{\lambda}f_{\lambda\mu}{}^{\varepsilon}+f_{\rho\mu}{}^{\lambda}%
f_{\lambda\upsilon}{}^{\varepsilon}\nonumber\\
&  =f_{\mu\nu}{}^{A}f_{A\rho}{}^{\varepsilon}+f_{\nu\rho}{}^{A}f_{A\mu}%
{}^{\varepsilon}+f_{\rho\mu}{}^{A}f_{A\upsilon}{}^{\varepsilon}+f_{\mu\nu}%
{}^{A+m}f_{(A+m)\rho}{}^{\varepsilon}\nonumber\\
&  +f_{\nu\rho}{}^{A+m}f_{(A+m)}{}^{\varepsilon}+f_{\rho\mu}{}^{A+m}%
f_{(A+m)\nu}{}^{\varepsilon}. \label{8}%
\end{align}

\begin{enumerate}
\item Case $\varepsilon=E$:

\begin{enumerate}
\item if $\mu=A$, $\nu=B$, $\rho=D$, we have
\begin{equation}
C_{AB}{}^{C}C_{CD}{}^{E}+C_{BD}{}^{C}C_{CA}{}^{E}+C_{DA}{}^{C}C_{CB}{}^{E}=0,
\label{9}%
\end{equation}

\item if $\mu=A$, $\nu=B$, $\rho=D+m,$ we have
\[
0=0,
\]

\item if $\mu=A$, $\nu=B+m$, $\rho=D+m$, we have
\begin{equation}
C_{BD}{}^{C}C_{CA}{}^{E}+C_{AB}{}^{C}C_{CD}{}^{E}+C_{DA}{}^{C}C_{CB}{}^{E}=0,
\label{10}%
\end{equation}

\item if $\mu=A+m$, $\nu=B+m$, $\rho=D+m$, we have
\[
0=0,
\]

\end{enumerate}

\item Case $\varepsilon=E+m$:

\begin{enumerate}
\item if $\mu=A$, $\nu=B$, $\rho=D$, we have
\[
0=0.
\]

\item if $\mu=A$, $\nu=B$, $\rho=D+m$, we have
\begin{equation}
C_{AB}{}^{C}C_{CD}{}^{E}+C_{BD}{}^{C}C_{CA}{}^{E}+C_{DA}{}^{C}C_{CB}{}^{E}=0,
\label{11}%
\end{equation}

\item if $\mu=A$, $\nu=B+m$, $\rho=D+m$, we have
\[
{0=0,}%
\]

\item if $\mu=A+m$, $\nu=B+m$, $\rho=D+m$, we have
\begin{equation}
C_{AB}{}^{C}C_{CD}{}^{E}+C_{BD}{}^{C}C_{CA}{}^{E}+C_{DA}{}^{C}C_{CB}{}^{E}=0.
\label{12}%
\end{equation}

\end{enumerate}
\end{enumerate}

This proves that, when the condition (\ref{cond1}) is imposed on an arbitrary
Lie algebra $Z_{4}\times\mathfrak{g}$, a new Lie algebra, which has half of
the generators of the $Z_{4}\times\mathfrak{g}$ algebra, is obtained.

\section{$H$\textbf{-Condition on }$Z_{2n}$\textbf{-expanded algebras }}

Now consider the generalization to the case of an arbitrary even-order cyclic
group. We have seen that the imposition of the $H$ condition (\ref{cond1}) on
the expanded algebra generates a new algebra. The generalization to the case
of an arbitrary cyclic group of even order can be carried out by rewriting
$Z_{2n}$, in the form%

\begin{align}
Z_{2n}  &  =\left\{  \lambda_{0},\ldots,\lambda_{n-1}\right\}  \cup\left\{
\lambda_{n},\ldots,\lambda_{2n-1}\right\}  ,\quad n\in N\label{13}\\
Z_{2n}  &  =\left\{  \lambda_{i}\right\}  \cup\left\{  \lambda_{i+n}\right\}
,\quad0\leq i\leq n-1
\end{align}
and then express the generators generated by the elements $\left\{
\lambda_{n},\ldots,\lambda_{2n-1}\right\}  $, which we will call "$greater$
$interval$", \ in terms of generators generated by elements $\left\{
\lambda_{0},\ldots,\lambda_{n-1}\right\}  $, which we will call "$minor$
$interval"$.

The generalization of $H$ condition(\ref{cond1}) is given by
\begin{equation}
T_{(A,i+n)}=-T_{(A,i)}. \label{eq:condgen}%
\end{equation}
Since the law of multiplication of $Z_{2n}$ is given by
\begin{equation}
\lambda_{\alpha}\lambda_{\beta}=\lambda_{\gamma\equiv\alpha+\beta\ \left(
\operatorname{mod}2n\right)  } \label{15}%
\end{equation}
we have that if we multiply equation (\ref{eq:condgen}) for arbitrary element
$\lambda_{\alpha}\in Z_{2n},$ we have (remember that $T_{(A,\alpha)}%
=\lambda_{\alpha}\times T_{A}$):

\begin{enumerate}
\item Case $\alpha=j$
\begin{align}
T_{(A,j+n+i\ \left(  \operatorname{mod}2n\right)  )}  &  =-T_{(A,j+i\ \left(
\operatorname{mod}2n\right)  )},\nonumber\\
T_{(A,(j+i)+n\ \left(  \operatorname{mod}2n\right)  )}  &
=-T_{(A,j+i\ \left(  \operatorname{mod}2n\right)  )}. \label{16}%
\end{align}
Because $\lambda_{j+i\ \left(  \operatorname{mod}2n\right)  }$ $\in Z_{2n}$,
we always obtain a non-trivial condition.

\item Case $\alpha=j+n$
\begin{align}
T_{(A,j+n+i+n\ \left(  \operatorname{mod}2n\right)  )}  &
=-T_{(A,j+n+i\ \left(  \operatorname{mod}2n\right)  )},\nonumber\\
T_{(A,j+i\ \left(  \operatorname{mod}2n\right)  )}  &
=-T_{(A,(j+i)+n\ \left(  \operatorname{mod}2n\right)  )}, \label{17}%
\end{align}
so that
\begin{equation}
T_{(A,(j+i)+n\ \left(  \operatorname{mod}2n\right)  )}=-T_{(A,j+i\ \left(
\operatorname{mod}2n\right)  )}. \label{18}%
\end{equation}

\end{enumerate}

From the closure property we have $\lambda_{j+i\ \left(  \operatorname{mod}%
2n\right)  }$ $\in Z_{2n}$, so in this case we also obtain a non-trivial condition.

The consistency condition (\ref{eq:condgen}) together with the previous
discussions allow us to establish the following theorem:

\textbf{Theorem 1}: Let $\mathfrak{g}=\left\{  T_{A}\right\}  $ be a Lie
algebra, with $A=1,\ldots,\dim\mathfrak{g}$ and let $Z_{2n}=\left\{
\lambda_{\alpha}\right\}  $ be a cyclic un group with $n\in N$ and
$\alpha=0,\ldots,2n-1$. The Algebra $\left\{  T_{(A,i)}\right\}  $,
$i=0,\ldots,n-1$, resulting from imposing the condition (\ref{eq:condgen}) on
the algebra expanded $Z_{2n}\times\mathfrak{g}$ is a Lie algebra whose
generators satisfy the following commutation relations
\begin{align}
\left[  T_{(A,i)},T_{(B,j)}\right]   &  =\left(  K_{ij}{}^{k}-K_{ij}{}%
^{k+n}\right)  C_{AB}{}^{C}T_{(C,k)},\quad\nonumber\\
A,B,C  &  =1,\ldots,\dim\mathfrak{g},\quad i,j,k=0,\cdot\cdot\cdot,n-1.
\label{19}%
\end{align}

\textbf{Proof}: The components $i,j=0,\cdot\cdot\cdot\cdot,n-1$ of the
commutations relations for the expanded algebra are given by
\begin{align}
\left[  T_{(A,i)},T_{(B,j)}\right]   &  =K_{ij}{}^{\gamma}C_{AB}{}%
^{C}T_{(C,\gamma)},\nonumber\\
&  =K_{ij}{}^{k}C_{AB}{}^{C}T_{(C,k)}+K_{ij}{}^{k+n}C_{AB}{}^{C}%
T_{(C,k+n)}\nonumber\\
&  =K_{ij}{}^{k}C_{AB}{}^{C}T_{(C,k)}-K_{ij}{}^{k+n}C_{AB}{}^{C}%
T_{(C,k)}\nonumber\\
&  =\left(  K_{ij}{}^{k}-K_{ij}{}^{k+n}\right)  C_{AB}{}^{C}T_{(C,k)},
\end{align}
where we have used the condition (\ref{eq:condgen}). The next step is to prove
that the structure constants
\begin{equation}
C_{\left(  A,i\right)  \left(  B,j\right)  }{}^{\left(  C,k\right)  }=\left(
K_{ij}{}^{k}-K_{ij}{}^{k+n}\right)  C_{AB}{}^{C} \label{20}%
\end{equation}
satisfy the antisymmetry property in its lower indices and the Jacobi
identity. From equation (\ref{20}) we have

\begin{proof}%
\begin{align}
C_{\left(  A,i\right)  \left(  B,j\right)  }{}^{\left(  C,k\right)  }  &
=\left(  K_{ij}{}^{k}-K_{ij}{}^{k+n}\right)  C_{AB}{}^{C}\nonumber\\
&  =\left(  K_{ji}{}^{k}-K_{ji}{}^{k+n}\right)  C_{AB}{}^{C}\nonumber\\
&  =-\left(  K_{ji}{}^{k}-K_{ji}{}^{k+n}\right)  C_{BA}{}^{C}\nonumber\\
&  =-C_{\left(  B,j\right)  \left(  A,i\right)  }{}^{\left(  C,k\right)  },
\label{21}%
\end{align}
which proves that the structure constants of the algebra $\left(  Z_{2n}%
\times\mathfrak{g}\right)  _{H}=\left\{  T_{(A,i)}\right\}  _{i=1}^{n-1}$ are
antisymmetric in their lower indices. To test the Jacobi identity consider the
following expression%

\begin{align}
&  C_{\left(  A,i\right)  \left(  B,j\right)  }{}^{\left(  C,k\right)
}C_{\left(  C,k\right)  \left(  D,l\right)  }{}^{\left(  E,m\right)
}+C_{\left(  B,j\right)  \left(  D,l\right)  }{}^{\left(  C,k\right)
}C_{\left(  C,k\right)  \left(  A,i\right)  }{}^{\left(  E,m\right)
}\nonumber\\
&  +C_{\left(  D,l\right)  \left(  A,i\right)  }{}^{\left(  C,k\right)
}C_{(C,k)\left(  B,j\right)  }{}^{\left(  E,m\right)  }\nonumber\\
&  =\left(  K_{ij}{}^{k}-K_{ij}{}^{k+n}\right)  C_{AB}{}^{C}\left(  K_{kl}%
{}^{m}-K_{kl}{}^{m+n}\right)  C_{CD}{}^{E}\nonumber\\
&  +\left(  K_{jl}{}^{k}-K_{jl}{}^{k+n}\right)  C_{BD}{}^{C}\left(  K_{ki}%
{}^{m}-K_{ki}{}^{m+n}\right)  C_{CA}{}^{E}\label{22}\\
&  +\left(  K_{li}{}^{k}-K_{li}{}^{k+n}\right)  C_{DA}{}^{C}\left(  K_{kj}%
{}^{m}-K_{kj}{}^{m+n}\right)  C_{CB}{}^{E}\nonumber
\end{align}
from where%
\begin{align}
&  C_{\left(  A,i\right)  \left(  B,j\right)  }{}^{\left(  C,k\right)
}C_{\left(  C,k\right)  \left(  D,l\right)  }{}^{\left(  E,m\right)
}+C_{\left(  B,j\right)  \left(  D,l\right)  }{}^{\left(  C,k\right)
}C_{\left(  C,k\right)  \left(  A,i\right)  }{}^{\left(  E,m\right)
}\nonumber\\
&  +C_{\left(  D,l\right)  \left(  A,i\right)  }{}^{\left(  C,k\right)
}C_{(C,k)\left(  B,j\right)  }{}^{\left(  E,m\right)  }\nonumber\\
&  =\left(  K_{ij}{}^{k}K_{kl}{}^{m}-K_{ij}{}^{k}K_{kl}{}^{m+n}-K_{ij}{}%
^{k+n}K_{kl}{}^{m}+K_{ij}{}^{k+n}K_{kl}{}^{m+n}\right)  C_{AB}{}^{C}C_{CD}%
{}^{E}\nonumber\\
&  +\left(  K_{jl}{}^{k}K_{ki}{}^{m}-K_{jl}{}^{k}K_{ki}{}^{m+n}-K_{jl}{}%
^{k+n}K_{ki}{}^{m}+K_{jl}{}^{k+n}K_{ki}{}^{m+n}\right)  C_{BD}{}^{C}C_{CA}%
{}^{E}\nonumber\\
&  +\left(  K_{li}{}^{k}K_{kj}{}^{m}-K_{li}{}^{k}K_{kj}{}^{m+n}-K_{li}{}%
^{k+n}K_{kj}{}^{m}+K_{li}{}^{k+n}K_{kj}{}^{m+n}\right)  C_{DA}{}^{C}C_{CB}%
{}^{E}. \label{23}%
\end{align}
Since the selectors satisfy the relation (see \cite{exp1}) $K_{\alpha
\beta\gamma}^{\text{ \ \ \ \ }\delta}=K_{\alpha\beta}{}^{\varepsilon
}K_{\varepsilon\gamma}^{\text{ \ }\delta}$ we have%
\begin{align}
K_{ijl}^{\text{ \ \ }m}  &  =K_{ij}{}^{k}K_{kl}{}^{m}+K_{ij}{}^{k+n}%
K_{k+n,l}{}^{m}\nonumber\\
K_{ijl}^{\text{ \ \ }m+n}  &  =K_{ij}{}^{k}K_{kl}{}^{m+n}+K_{ij}{}%
^{k+n}K_{k+n,l}{}^{m+n}%
\end{align}
so that
\begin{align}
&  C_{\left(  A,i\right)  \left(  B,j\right)  }{}^{\left(  C,k\right)
}C_{\left(  C,k\right)  \left(  D,l\right)  }{}^{\left(  E,m\right)
}+C_{\left(  B,j\right)  \left(  D,l\right)  }{}^{\left(  C,k\right)
}C_{\left(  C,k\right)  \left(  A,i\right)  }{}^{\left(  E,m\right)
}\nonumber\\
&  +C_{\left(  D,l\right)  \left(  A,i\right)  }{}^{\left(  C,k\right)
}C_{(C,k)\left(  B,j\right)  }{}^{\left(  E,m\right)  }\nonumber\\
&  =\left(  K_{ijl}^{\text{ \ \ }m}-K_{ij}{}^{k+n}K_{k+n,l}{}^{m}%
-K_{ijl}^{\text{ \ \ }m+n}+K_{ij}{}^{k+n}K_{k+n,l}{}^{m+n}\right. \nonumber\\
&  \left.  -K_{ij}{}^{k+n}K_{kl}{}^{m}+K_{ij}{}^{k+n}K_{kl}{}^{m+n}\right)
C_{AB}{}^{C}C_{CD}{}^{E}\nonumber\\
&  +\left(  K_{jli}^{\text{ \ \ }m}-K_{jl}{}^{k+n}K_{k+n,i}{}^{m}%
-K_{jli}^{\text{ \ \ }m+n}+K_{jl}{}^{k+n}K_{k+n,i}{}^{m+n}\right. \nonumber\\
&  \left.  -K_{jl}{}^{k+n}K_{ki}{}^{m}+K_{jl}{}^{k+n}K_{ki}{}^{m+n}\right)
C_{BD}{}^{C}C_{CA}{}^{E}\nonumber\\
&  +\left(  K_{lij}^{\text{ \ \ }m}-K_{li}{}^{k+n}K_{k+n,j}{}^{m}%
-K_{lij}^{\text{ \ \ }m+n}+K_{li}{}^{k+n}K_{k+n,j}{}^{m+n}\right. \nonumber\\
&  \left.  -K_{li}{}^{k+n}K_{kj}{}^{m}+K_{li}{}^{k+n}K_{kj}{}^{m+n}\right)
C_{DA}{}^{C}C_{CB}{}^{E}.
\end{align}
Taking into account that $K_{\alpha\beta\gamma}^{\text{ \ \ \ \ }\delta
}=K_{\beta\gamma\alpha}^{\text{ \ \ \ \ }\delta}=K_{\gamma\alpha\beta}^{\text{
\ \ \ \ }\delta}$ (see \cite{exp1}) and that the algebra $\mathfrak{g}$
satisfies the identity Jacobi, we have%
\begin{align}
&  C_{\left(  A,i\right)  \left(  B,j\right)  }{}^{\left(  C,k\right)
}C_{\left(  C,k\right)  \left(  D,l\right)  }{}^{\left(  E,m\right)
}+C_{\left(  B,j\right)  \left(  D,l\right)  }{}^{\left(  C,k\right)
}C_{\left(  C,k\right)  \left(  A,i\right)  }{}^{\left(  E,m\right)
}\nonumber\\
&  +C_{\left(  D,l\right)  \left(  A,i\right)  }{}^{\left(  C,k\right)
}C_{(C,k)\left(  B,j\right)  }{}^{\left(  E,m\right)  }\nonumber\\
&  =K_{ij}{}^{k+n}\left(  -K_{k+n,l}^{\text{ \ \ \ \ \ }m}+K_{k+n,l}{}%
^{m+n}-K_{kl}^{\text{ \ \ }m}+K_{kl}{}^{m+n}\right)  C_{AB}{}^{C}C_{CD}{}%
^{E}\nonumber\\
&  +K_{jl}{}^{k+n}\left(  -K_{k+n,i}^{\text{ \ \ \ \ \ }m}+K_{k+n,i}{}%
^{m+n}-K_{ki}{}^{m}+K_{ki}{}^{m+n}\right)  C_{BD}{}^{C}C_{CA}{}^{E}\nonumber\\
&  +\left(  -K_{k+n,j}^{\text{ \ \ \ \ \ }m}+K_{k+n,j}{}^{m+n}-K_{kj}{}%
^{m}+K_{kj}{}^{m+n}\right)  C_{DA}{}^{C}C_{CB}{}^{E}%
\end{align}
Since $K_{k+n,l}^{\text{ \ \ \ \ \ }m}=K_{kl}{}^{m+n}$ and $K_{k+n,l}{}%
^{m+n}=K_{kl}{}^{m}$ (see appendix) we have each parenthesis of the above
equation vanishes, proving the Jacobi identity.
\end{proof}

\subsection{\textbf{Lorentz algebra from }$so(3)$\textbf{ algebra}}

Using the S-expansion procedure with $S=Z_{4}=\left\{  \lambda_{0},\lambda
_{1},\lambda_{2},\lambda_{3}\right\}  $ is found that the generators
$J_{(i,\alpha)}=\lambda_{\alpha}J_{i}$ of the $Z_{4}\times\mathfrak{so}(3)$
algebra satisfy the following commutation relations \
\begin{equation}%
\begin{array}
[c]{ll}%
\left[  J_{(i,0)},J_{(j,0)}\right]  =\varepsilon_{ij}^{k}J_{(k,0)},\quad &
\left[  J_{(i,0)},J_{(j,1)}\right]  =\varepsilon_{ij}^{k}J_{(k,1)},\\
\left[  J_{(i,0)},J_{(j,2)}\right]  =\varepsilon_{ij}^{k}J_{(k,2)},\quad &
\left[  J_{(i,0)},J_{(j,3)}\right]  =\varepsilon_{ij}^{k}J_{(k,3)},\\
\left[  J_{(i,1)},J_{(j,1)}\right]  =\varepsilon_{ij}^{k}J_{(k,2)},\quad &
\left[  J_{(i,1)},J_{(j,2)}\right]  =\varepsilon_{ij}^{k}J_{(k,3)},\\
\left[  J_{(i,1)},J_{(j,3)}\right]  =\varepsilon_{ij}^{k}J_{(k,0)},\quad &
\left[  J_{(i,2)},J_{(j,2)}\right]  =\varepsilon_{ij}^{k}J_{(k,0)},\\
\left[  J_{(i,2)},J_{(j,3)}\right]  =\varepsilon_{ij}^{k}J_{(k,1)},\quad &
\left[  J_{(i,3)},J_{(j,3)}\right]  =\varepsilon_{ij}^{k}J_{(k,2)},
\end{array}
\label{27}%
\end{equation}
Imposing the condition
\begin{equation}
J_{(i,2)}=-J_{(i,0)},\quad J_{(i,3)}=-J_{(i,1)}%
\end{equation}
on the expanded algebra $Z_{4}\times\mathfrak{so}(3)$ we obtain
\begin{equation}
\mathfrak{so}(3,1)\simeq\left(  Z_{4}\times\mathfrak{so}(3)\right)  _{H}.
\label{eq:alglorentzh}%
\end{equation}

\subsection{\textbf{Dimension of a }$S_{H}$\textbf{-expanded algebra}}

From \cite{exp1} we know that if $\mathfrak{g}$ is a Lie algebra of dimension
$\dim\mathfrak{g}$, then the dimension of $S$-expanded algebra $S\times
\mathfrak{g}$ is given by $\left\vert S\right\vert \dim\mathfrak{g}$. In the
case that $S=Z_{2n}$ we have
\begin{equation}
\dim\left(  Z_{2n}\times\mathfrak{g}\right)  _{H}=n\dim\mathfrak{g}.
\label{eq:dimh}%
\end{equation}
In fact, since the dimension of the algebra $Z_{2n}\times\mathfrak{g}$ is
$2n\dim\mathfrak{g}$, we have that if we impose the $H$ condition on
$Z_{2n}\times\mathfrak{g}$, then it follows that the number of generators
$T_{(A,i)}$ is halved, i.e., $i=0,1,\cdot\cdot\cdot,n-1.$ Therefore the
dimension of the $S_{H}$-expanded algebra is given by $n\dim\mathfrak{g}$.

\subsection{\textbf{Construction with the greater interval}}

The construction of an algebra $S_{H}$-expanded is performed with generators
obtained using only elements minor interval. It is of interest to analyze the
characteristics that has a algebra constructed from generators obtained using
only elements greater interval.

Evaluating the components $i+n$ of the commutation relations for $Z_{2n}%
$-expanded algebra we have
\begin{align}
\left[  T_{(A,i+n)},T_{(B,j+n)}\right]   &  =K_{i+n,j+n}^{\gamma}C_{AB}{}%
^{C}T_{(C,\gamma)},\nonumber\\
&  =K_{i+n,j+n}^{k}C_{AB}{}^{C}T_{(C,k)}+K_{i+n,j+n}^{k+n}C_{AB}{}%
^{C}T_{(C,k+n)},\nonumber\\
\quad &  =-K_{i+n,j+n}^{k}C_{AB}{}^{C}T_{(C,k+n)}+K_{i+n,j+n}^{k+n}C_{AB}%
{}^{C}T_{(C,k+n)},\label{28}\\
\quad &  =-\left(  K_{ij}^{k}-K_{ij}^{k+n}\right)  C_{AB}{}^{C}T_{(C,k+n)}%
.\nonumber
\end{align}
From (\ref{28}) can see that when changing base $T_{(A,i+n)}^{\prime
}=-T_{(A,i+n)}$ we obtain an algebra isomorphic to the algebra $S_{H}%
$-expanded, generated by generators $T_{(A,i)}$.

\subsection{$S_{H}$\textbf{-expansion in the case that }$S=Z_{2}$}

If $n=1$, the $S_{H}$-expansion corresponds to the trivial case, i.e.,
\begin{equation}
(Z_{2}\times\mathfrak{g})_{H}\simeq\mathfrak{g}. \label{eq:z2hred}%
\end{equation}
In fact, from the previous theorem we can see that the commutation relations
$(Z_{2}\times\mathfrak{g})_{H}$ are given by
\begin{align}
\left[  T_{(A,0)},T_{(B,0)}\right]   &  =\left(  K_{00}{}^{k}-K_{00}{}%
^{k+1}\right)  C_{AB}{}^{C}T_{(C,k)},\nonumber\\
&  =\left(  K_{00}{}^{0}-K_{00}{}^{1}\right)  C_{AB}{}^{C}T_{(C,0)}%
,\nonumber\\
&  =C_{AB}{}^{C}T_{(C,0)},
\end{align}
which proves that the obtained algebra is an algebra isomorphic to
$\mathfrak{g}$.

\subsection{\textbf{The Klein group}}

The Klein group $D_{4}$ corresponds to the direct product $Z_{2}\times Z_{2}$
(see Appendix 1). Taking into account that the $S^{\prime}$-expansion, of a
$S$-expanded algebra, corresponds to an $(S^{\prime}\times S)$-expansion, we
have
\begin{equation}
D_{4}\times\mathfrak{g}\simeq Z_{2}\times\left(  Z_{2}\times\mathfrak{g}%
\right)  .
\end{equation}
From (\ref{eq:z2hred}) we know that applying the $H$ condition on an algebra
$Z_{2}$-expanded, we get an algebra isomorphic to the original algebra. This
means that
\begin{equation}
\left(  D_{4}\times\mathfrak{g}\right)  _{H}\simeq\left(  Z_{2}\times\left(
Z_{2}\times\mathfrak{g}\right)  \right)  _{H}\simeq Z_{2}\times\mathfrak{g}.
\end{equation}

It is straightforward to prove that the imposition of the condition
$T_{(A,2)}=-T_{(A,0)}$ on $D_{4}\times\mathfrak{g}$, leads to a Lie algebra
$Z_{2}$-expanded. Since the $S_{H}$-expansion procedure has been so far
defined only for cyclic groups, this result suggests the possibility of
generalizing the $S_{H}$-expansion procedure to the case of a larger family of
abelian groups.

Since a $Z_{2}$-expansion (which corresponds to a "$\left(  D_{4}\right)
_{H}$-expansion") of $\mathfrak{so}(3)$ algebra leads to algebra
$\mathfrak{so}(4)$ and that a $\left(  Z_{4}\right)  _{H}$-expansion of the
$\mathfrak{so}(3)$ algebra leads to the algebra $\mathfrak{so}(3,1)$ we can
write
\begin{equation}
\mathfrak{so}(3,1)\simeq\left(  Z_{4}\times\mathfrak{so}(3)\right)  _{H}%
,\quad\mathfrak{so}(4)\simeq\left(  D_{4}\times\mathfrak{so}(3)\right)  _{H},
\label{eq:comparacion}%
\end{equation}
where $Z_{4}$ and $D_{4}$ are the only groups of order $4$ that there, up to isomorphism.

\section{\textbf{Invariant tensors and\ dual formulation of }$S_{H}%
$\textbf{-expansion procedure}}

In this section, we obtain the invariant tensors corresponding to $S_{H}%
$-expanded algebras. The dual formulation of $S_{H}$-expansion procedure is
also considered.

\subsection{\textbf{ Invariant Tensors}}

In ref. \cite{exp1} was proved that the invariant tensors, corresponding to an
S-expanded algebra, can be obtained from the invariant tensors of the original
algebra. We now prove that this result remains valid in the case of $S_{H}%
$--expanded algebras .

\textbf{Theorem 2:} Let $\left\langle T_{A_{1}},\ldots,T_{A_{r}}\right\rangle
$ be an invariant tensor for a Lie algebra $\mathfrak{g}$ of basis $\left\{
T_{A}\right\}  $. Then the expression
\begin{equation}
\left\langle T_{(A_{1},i_{1})},\ldots,T_{(A_{r},i_{r})}\right\rangle
=\alpha_{\gamma}K_{i_{1}\ldots i_{r}}^{\text{ \ \ \ \ \ \ \ }\gamma
}\left\langle T_{A_{1}},\ldots,T_{A_{r}}\right\rangle , \label{eq:tensinvh}%
\end{equation}
where $K_{\alpha_{1}\ldots\alpha_{r}}^{\text{ \ \ \ \ \ \ \ \ \ }\gamma}$ is
the $r$-selector for $Z_{2n},$ corresponds to an invariant tensor for the
$S_{H}$-expanded algebra\textbf{ }$(Z_{2n}\times\mathfrak{g})_{H}$, with
$i_{1},\ldots,i_{r}\in\left\{  0,\ldots,n-1\right\}  $ and $0\leq i_{k}\leq
n-1.$

\textbf{Proof: }The invariance condition for $\left\langle T_{A_{1}}%
,\ldots,T_{A_{r}}\right\rangle $ under $\mathfrak{g}$ reads
\begin{equation}
\sum_{p=1}^{r}X_{A_{0}\cdots A_{r}}^{\left(  p\right)  }=0, \label{147}%
\end{equation}
where%
\begin{equation}
X_{A_{0}\cdots A_{r}}^{\left(  p\right)  }=C_{A_{0}A_{p}}^{\text{\qquad}%
B}\left\langle \boldsymbol{T}_{A_{1}}\cdots\boldsymbol{T}_{A_{p-1}%
}\boldsymbol{T}_{B}\boldsymbol{T}_{A_{p+1}}\cdots\boldsymbol{T}_{A_{r}%
}\right\rangle . \label{InvTensorS-Exp}%
\end{equation}

Define now
\begin{align*}
X_{\left(  A_{0},\alpha_{0}\right)  \cdots\left(  A_{r},\alpha_{r}\right)
}^{\left(  p\right)  }  &  =C_{\left(  A_{0},i_{0}\right)  \left(  A_{p}%
,i_{p}\right)  }^{\text{\qquad\qquad\qquad}\left(  B,\beta\right)
}\left\langle \boldsymbol{T}_{\left(  A_{1},i_{1}\right)  }\cdots
\boldsymbol{T}_{\left(  A_{p-1},i_{p-1}\right)  }\boldsymbol{T}_{\left(
B,\beta\right)  }\boldsymbol{T}_{\left(  A_{p+1},i_{p+1}\right)  }%
\cdots\boldsymbol{T}_{\left(  A_{r},\alpha_{r}\right)  }\right\rangle \\
&  =C_{\left(  A_{0},i_{0}\right)  \left(  A_{p},i_{p}\right)  }%
^{\text{\qquad\qquad\qquad}\left(  B,j\right)  }\left\langle \boldsymbol{T}%
_{\left(  A_{1},i_{1}\right)  }\cdots\boldsymbol{T}_{\left(  A_{p-1}%
,i_{p-1}\right)  }\boldsymbol{T}_{\left(  B,j\right)  }\boldsymbol{T}_{\left(
A_{p+1},i_{p+1}\right)  }\cdots\boldsymbol{T}_{\left(  A_{r},\alpha
_{r}\right)  }\right\rangle \\
&  +C_{\left(  A_{0},i_{0}\right)  \left(  A_{p},i_{p}\right)  }%
^{\text{\qquad\qquad\qquad}\left(  B,j+n\right)  }\left\langle \boldsymbol{T}%
_{\left(  A_{1},i_{1}\right)  }\cdots\boldsymbol{T}_{\left(  A_{p-1}%
,i_{p-1}\right)  }\boldsymbol{T}_{\left(  B,j+n\right)  }\boldsymbol{T}%
_{\left(  A_{p+1},i_{p+1}\right)  }\cdots\boldsymbol{T}_{\left(  A_{r}%
,\alpha_{r}\right)  }\right\rangle .
\end{align*}
Imposing the $H$-condition, we have%
\begin{align*}
X_{\left(  A_{0},\alpha_{0}\right)  \cdots\left(  A_{r},\alpha_{r}\right)
}^{\left(  p\right)  }  &  =C_{\left(  A_{0},i_{0}\right)  \left(  A_{p}%
,i_{p}\right)  }^{\text{\qquad\qquad\qquad}\left(  B,j\right)  }\left\langle
\boldsymbol{T}_{\left(  A_{1},i_{1}\right)  }\cdots\boldsymbol{T}_{\left(
A_{p-1},i_{p-1}\right)  }\boldsymbol{T}_{\left(  B,j\right)  }\boldsymbol{T}%
_{\left(  A_{p+1},i_{p+1}\right)  }\cdots\boldsymbol{T}_{\left(  A_{r}%
,\alpha_{r}\right)  }\right\rangle \\
&  -C_{\left(  A_{0},i_{0}\right)  \left(  A_{p},i_{p}\right)  }%
^{\text{\qquad\qquad\qquad}\left(  B,j+n\right)  }\left\langle \boldsymbol{T}%
_{\left(  A_{1},i_{1}\right)  }\cdots\boldsymbol{T}_{\left(  A_{p-1}%
,i_{p-1}\right)  }\boldsymbol{T}_{\left(  B,j\right)  }\boldsymbol{T}_{\left(
A_{p+1},i_{p+1}\right)  }\cdots\boldsymbol{T}_{\left(  A_{r},\alpha
_{r}\right)  }\right\rangle
\end{align*}
and replacing the expressions~$C_{(A,\alpha)\left(  B,\beta\right)  }^{\text{
\ \ \ \ \ \ \ \ \ \ \ \ \ }\left(  C,\gamma\right)  }=K_{\alpha\beta}^{\text{
\ \ \ }\gamma}C_{AB}^{\text{ \ \ \ }C}$ for the $S$-expansion structure
constants and~(\ref{InvTensorS-Exp}) for $\left\langle \boldsymbol{T}_{\left(
A_{1},\alpha_{1}\right)  }\cdots\boldsymbol{T}_{\left(  A_{n},\alpha
_{n}\right)  }\right\rangle $, we get%
\begin{align*}
X_{\left(  A_{0},i_{0}\right)  \cdots\left(  A_{n},i_{n}\right)  }^{\left(
p\right)  }  &  =\left(  K_{i_{0}i_{p}}^{\text{ \ \ \ }j}-K_{i_{0}i_{p}%
}^{\text{ \ \ \ \ }j+n}\right)  C_{A_{0}A_{p}}^{\text{\qquad}B}\alpha_{\gamma
}K_{i_{1}\cdot\cdot\cdot\cdot i_{p-1}ji_{p+1}\cdot\cdot\cdot i_{r}}^{\text{
\ \ \ \ \ \ \ \ \ \ \ \ \ \ \ \ \ \ \ \ \ \ \ \ \ \ }\gamma}\left\langle
\boldsymbol{T}_{A_{1}}\cdots\boldsymbol{T}_{A_{p-1}}\boldsymbol{T}%
_{B}\boldsymbol{T}_{A_{p+1}}\cdots\boldsymbol{T}_{A_{r}}\right\rangle \\
X_{\left(  A_{0},i_{0}\right)  \cdots\left(  A_{n},i_{n}\right)  }^{\left(
p\right)  }  &  =\alpha_{\gamma}\left(  K_{i_{0}i_{p}}^{\text{ \ \ \ }%
j}-K_{i_{0}i_{p}}^{\text{ \ \ \ \ }j+n}\right)  K_{i_{1}\cdot\cdot\cdot\cdot
i_{p-1}ji_{p+1}\cdot\cdot\cdot i_{r}}^{\text{
\ \ \ \ \ \ \ \ \ \ \ \ \ \ \ \ \ \ \ \ \ \ \ \ \ \ }\gamma}X_{A_{0}\cdots
A_{r}}^{\left(  p\right)  }%
\end{align*}
and from~(\ref{147}) one readily concludes that
\begin{equation}
\sum_{p=1}^{n}X_{\left(  A_{0},\alpha_{0}\right)  \cdots\left(  A_{n}%
,\alpha_{n}\right)  }^{\left(  p\right)  }=0.
\end{equation}

Therefore, $\left\langle \boldsymbol{T}_{\left(  A_{1},i_{1}\right)  }%
\cdots\boldsymbol{T}_{\left(  A_{n},i_{n}\right)  }\right\rangle
=\alpha_{\gamma}K_{i_{1}\cdots i_{n}}^{\qquad\;\gamma}\left\langle
\boldsymbol{T}_{A_{1}}\cdots\boldsymbol{T}_{A_{n}}\right\rangle $ is an
invariant tensor for $\mathfrak{G}=Z_{2n}\otimes\mathfrak{g}$.

\section{\textbf{Dual formulation of the }$S_{H}$\textbf{-expansion
procedure}}

In Ref. \cite{exp3} was found a dual formulation of the Lie algebra
$S$-expansion procedure, which is based on the dual picture of a Lie algebra
given by the Maurer-Cartan forms. \ In this section we considere the dual
formulation of the $S_{H}$-expansion procedure.

\textbf{Theorem 3}: Let $\omega^{(A,\gamma)}$ be the Maurer-Cartan forms for
the expanded Lie algebra $Z_{2n}\times\mathfrak{g}$, which satisfy the
Maurer-Cartan equations%
\begin{equation}
d\omega^{(C,\gamma)}+\frac{1}{2}K_{\alpha\beta}^{\text{ \ \ }\gamma}%
C_{AB}^{\text{ \ \ }C}\omega^{(A,\alpha)}\omega^{(B,\beta)}=0 \label{dual1}%
\end{equation}
Imposing the $H$-condition
\begin{equation}
\omega^{(A,i+n)}\overset{!}{=}-\omega^{(A,i)};\text{ }i=1,\ldots,n-1
\label{eq:condgendual}%
\end{equation}
on the expanded Lie algebra $Z_{2n}\times\mathfrak{g}$ we obtain the
Maurer-Cartan equations for the $S_{H}$-expanded Lie algebra $\left(
Z_{2n}\times\mathfrak{g}\right)  _{H}$.

\textbf{Proof: \ }From equation (\ref{dual1}) we can see that the components
$k=0,\cdot\cdot\cdot\cdot,n-1$ of the Maurer-Cartan equations for the
$S$-expanded Lie algebra are%
\begin{equation}
d\omega^{(C,k)}+\frac{1}{2}K_{\alpha\beta}^{\text{ \ \ }k}C_{AB}^{\text{
\ \ }C}\omega^{(A,\alpha)}\omega^{(B,\beta)}=0. \label{dual2}%
\end{equation}

Since%
\begin{align*}
K_{\alpha\beta}^{\text{ \ \ }k}\omega^{(A,\alpha)}\omega^{(B,\beta)}  &
=K_{i\beta}^{\text{ \ \ }k}\omega^{(A,i)}\omega^{(B,\beta)}+K_{i+n,\beta
}^{\text{ \ \ }k}\omega^{(A,i+n)}\omega^{(B,\beta)}\\
&  =K_{ij}^{\text{ \ \ }k}\omega^{(A,i)}\omega^{(B,j)}+K_{i,j+n}^{\text{
\ \ }k}\omega^{(A,i)}\omega^{(B,j+n)}\\
&  +K_{i+n,j}^{\text{ \ \ }k}\omega^{(A,i+n)}\omega^{(B,j)}+K_{i+n,j+n}%
^{\text{ \ \ }k}\omega^{(A,i+n)}\omega^{(B,j+n)}%
\end{align*}

Imposing the $H$-condition (\ref{eq:condgendual}), we have%
\[
K_{\alpha\beta}^{\text{ \ \ }k}\omega^{(A,\alpha)}\omega^{(B,\beta)}=\left(
K_{ij}^{\text{ \ \ }k}-K_{i,j+n}^{\text{ \ \ \ \ \ }k}-K_{i+n,j}^{\text{
\ \ \ \ \ \ }k}+K_{i+n,j+n}^{\text{ \ \ \ \ \ \ \ \ \ }k}\right)
\omega^{(A,i)}\omega^{(B,j)}.
\]

So that Eq. (\ref{dual2}) takes the form%
\[
d\omega^{(C,k)}+\frac{1}{2}\left(  K_{ij}^{\text{ \ \ }k}-K_{i,j+n}^{\text{
\ \ \ \ \ }k}-K_{i+n,j}^{\text{ \ \ \ \ \ }k}+K_{i+n,j+n}^{\text{
\ \ \ \ \ \ \ \ \ }k}\right)  C_{AB}^{\text{ \ \ }C}\omega^{(A,i)}%
\omega^{(B,j)}=0.
\]

Since the selectors of $S=Z_{2n}$ satisfy the equations $K_{i+n,j}^{\text{
\ \ \ \ \ }k}=K_{ij}^{\text{ \ \ }k+n}$ ;$K_{i+n,j+n}^{\text{
\ \ \ \ \ \ \ \ \ }\gamma}=K_{ij}^{\text{ \ \ }\gamma}$ ; $K_{i,j+n}^{\text{
\ \ \ \ \ }k}=K_{i+n,j}^{\text{ \ \ \ \ \ }k}$ we have%
\[
d\omega^{(C,k)}+\frac{1}{2}\left[  2\left(  K_{ij}^{\text{ \ \ }k}%
-K_{ij}^{\text{ \ \ }k+n}\right)  \right]  C_{AB}^{\text{ \ \ }C}%
\omega^{(A,i)}\omega^{(B,j)}=0.
\]

Therefore $\omega^{(A,i)}$ are the Maurer-Cartan forms for a Lie algebra,
which is isomorphic to $\left(  Z_{2n}\times\mathfrak{g}\right)  _{H}$ whose
structure constants are given by
\[
C_{(A,i)\left(  B,i\right)  }^{\text{ \ \ }\left(  C,\gamma\right)  }=\left[
2\left(  K_{ij}^{\text{ \ \ }k}-K_{ij}^{\text{ \ \ }k+n}\right)  \right]
C_{AB}^{\text{ \ \ }C}.
\]

\section{\textbf{General Relativity with cosmological constant from
Chern-Simons gravity}}

According to the principles of general relativity (GR), the spacetime is a
dynamical object which has independent degrees of freedom, and is governed by
dynamical equations, namely the Einstein field equations. This means that in
GR the geometry is dynamically determined. Therefore, the construction of a
gauge theory of gravity requires an action that does not consider a fixed
space-time background. A five dimensional action for gravity fulfilling these
conditions is the five-dimensional Chern--Simons AdS gravity action, which can
be written as%

\begin{equation}
L_{\mathrm{AdS}}^{\left(  5\right)  }=\kappa\left(  \frac{1}{5l^{5}}%
\epsilon_{a_{1}\cdots a_{5}}e^{a_{1}}\cdots e^{a_{5}}+\frac{2}{3l^{3}}%
\epsilon_{a_{1}\cdots a_{5}}R^{a_{1}a_{2}}e^{a_{3}}\cdots e^{a_{5}}+\frac
{1}{l}\epsilon_{a_{1}\cdots a_{5}}R^{a_{1}a_{2}}R^{a_{3}a_{4}}e^{a_{5}%
}\right)  ,
\end{equation}
where $e^{a}$ corresponds to the 1-form \emph{vielbein}, and $R^{ab}%
=d\omega^{ab}+\omega_{\text{ }c}^{a}\omega^{cb}$ to the Riemann curvature in
the first order formalism \cite{cham1}, \cite{cham2}, \cite{zan}.

If Chern-Simons theories are the appropriate gauge-theories to provide a
framework for the gravitational interaction, then these theories must satisfy
the correspondence principle, namely they must be related to General Relativity.

In ref. \cite{standardg} was shown that the standard, five-dimensional General
Relativity, $without$ $a$ $cosmological$ $term$, can be obtained from
Chern-Simons gravity theory for a certain Lie algebra $\mathcal{B}$. The
Chern-Simons Lagrangian is built from a $\mathcal{B}$-valued, one-form gauge
connection $A$ which depends on a scale parameter $l$ which can be interpreted
as a coupling constant that characterizes different regimes within the theory.
The $\mathcal{B}$ algebra, on the other hand, is obtained from the $AdS$
algebra and a particular semigroup $S$ by means of the S-expansion procedure
introduced in refs. \cite{exp1}, \cite{exp3}. The field content induced by
$\mathcal{B}$ includes the vielbein $e^{a}$, the spin connection $\omega^{ab}$
and two extra bosonic fields $h^{a}$ and $k^{ab}$.

The five dimensional Chern-Simons Lagrangian for the $\mathcal{B}$ algebra is
given by \cite{standardg}:%

\begin{equation}
L_{\mathrm{ChS}}^{(5)}=\alpha_{1}l^{2}\varepsilon_{abcde}R^{ab}R^{cd}%
e^{e}+\alpha_{3}\varepsilon_{abcde}\left(  \frac{2}{3}R^{ab}e^{c}e^{d}%
e^{e}+2l^{2}k^{ab}R^{cd}T^{\text{ }e}+l^{2}R^{ab}R^{cd}h^{e}\right)  ,
\label{26}%
\end{equation}
where we can see that $(i)$ if one identifies the field $e^{a}$ with the
vielbein, the system consists of the Einstein-Hilbert action, without a
cosmological, constant plus nonminimally coupled matter fields given by
$h^{a}$ and $k^{ab};$ $(ii)$ it is possible to recover the odd-dimensional
Einstein gravity theory from a Chern-Simons gravity theory in the limit where
the coupling constant $l$ equals to zero while keeping the effective Newton's
constant fixed.

In this section it is shown that the five-dimensional Einstein-Hilbert action
$with$ $a$ $cosmological$ $term$ can be obtained from a Chern-Simons gravity
action, invariant under a Lie algebra obtained by $S_{H}$-expansion of the
$AdS$ algebra.

\subsection{$S_{H}$\textbf{-expansion of AdS}$_{5}$\textbf{ algebra}}

From Ref.\cite{standardg} we know that the $\mathfrak{B}_{5}$-algebra can be
obtained by $S_{E}^{(3)}$-expansion resonant and reduced of the $AdS$ algebra,
i.e.,
\begin{equation}
\mathfrak{B}_{5}=\left(  S_{E}^{(3)}\times AdS_{5}\right)  _{R,0_{S}},
\label{29}%
\end{equation}
where $R,0_{S}$ denotes resonance followed by a $0_{S}$-reduction. This
algebra has $30$ generators which are denoted by $J_{ab}$, $Z_{ab}$, $P_{a}$
and $Z_{a}$ with $a,b=1,\ldots,5$. \ Now we consider a new algebra, which has
also $30$ generators, and that can be obtained by $S_{H}$-expansion of the
$AdS_{5}$ algebra, which will be denoted by $\mathfrak{C}_{5},$ where
\begin{equation}
\mathfrak{C}_{5}=\left(  Z_{4}\times AdS_{5}\right)  _{H}. \label{30}%
\end{equation}

The generators of $AdS_{5}$ algebra satisfy the following commutation
relation
\[
\left[  \tilde{J}_{ab},\tilde{J}_{cd}\right]  =f_{ab,cd}^{\text{
\ \ \ \ \ \ \ }ef}\tilde{J}_{ef},\quad\left[  \tilde{J}_{ab},\tilde{P}%
_{c}\right]  =f_{ab,c}^{\text{ \ \ \ \ \ }d}\tilde{P}_{d},\quad\left[
\tilde{P}_{a},\tilde{P}_{b}\right]  =\tilde{J}_{ab},
\]
where%
\[
f_{ab,cd}^{\text{ \ \ \ \ \ \ }ef}=-\frac{1}{2}\left\{  \eta_{ac}\left(
\delta_{b}^{e}\delta_{d}^{f}-\delta_{d}^{e}\delta_{b}^{f}\right)  +\eta
_{bd}\left(  \delta_{a}^{e}\delta_{c}^{f}-\delta_{c}^{e}\delta_{a}^{f}\right)
+\eta_{cb}\left(  \delta_{d}^{e}\delta_{a}^{f}-\delta_{a}^{e}\delta_{d}%
^{f}\right)  +\eta_{da}\left(  \delta_{c}^{e}\delta_{b}^{f}-\delta_{b}%
^{e}\delta_{c}^{f}\right)  \right\}
\]%
\[
f_{ab,c}^{\text{ \ \ \ \ \ }e}=-\left(  \eta_{ac}\delta_{b}^{e}-\eta
_{bc}\delta_{a}^{e}\right)
\]

The corresponding commutation relation of the generators of the $S_{H}%
$-expanded $AdS_{5}$ algebra, which will be denoted by $\mathfrak{C}_{5}$, are
given by
\[%
\begin{array}
[c]{ll}%
\left[  \tilde{J}_{(ab,0)},\tilde{J}_{(cd,0)}\right]  =f_{ab,cd}^{\text{
\ \ \ \ \ \ \ }ef}\tilde{J}_{(ef,0)}, & \left[  \tilde{J}_{(ab,1)},\tilde
{P}_{(c,0)}\right]  =f_{ab,c}^{\text{ \ \ \ \ }d}\tilde{P}_{(d,1)},\\
\left[  \tilde{J}_{(ab,0)},\tilde{J}_{(cd,1)}\right]  =f_{ab,cd}^{\text{
\ \ \ \ \ \ \ }ef}\tilde{J}_{(ef,1)}, & \left[  \tilde{J}_{(ab,1)},\tilde
{P}_{(c,1)}\right]  =-f_{ab,c}^{\text{ \ \ \ \ }d}\tilde{P}_{(d,0)},\\
\left[  \tilde{J}_{(ab,1)},\tilde{J}_{(cd,1)}\right]  =-f_{ab,cd}^{\text{
\ \ \ \ \ \ }ef}\tilde{J}_{(ef,0)}, & \left[  \tilde{P}_{(a,0)},\tilde
{P}_{(b,0)}\right]  =\tilde{J}_{(ab,0)},\\
\left[  \tilde{J}_{(ab,0)},\tilde{P}_{(c,0)}\right]  =f_{ab,c}^{\text{
\ \ \ \ }d}\tilde{P}_{(d,0)}, & \left[  \tilde{P}_{(a,0)},\tilde{P}%
_{(b,1)}\right]  =\tilde{J}_{(ab,1)},\\
\left[  \tilde{J}_{(ab,0)},\tilde{P}_{(c,1)}\right]  =f_{ab,c}^{\text{
\ \ \ \ }d}\tilde{P}_{(d,1)}, & \left[  \tilde{P}_{(a,1)},\tilde{P}%
_{(b,1)}\right]  =-\tilde{J}_{(ab,0)},
\end{array}
\]
The identification $J_{(ab,0)}\equiv J_{ab}$, $J_{(ab,1)}\equiv Z_{ab}$,
$P_{(a,0)}\equiv P_{a}$ and $P_{(a,1)}\equiv Z_{a}$, leads
\begin{equation}%
\begin{array}
[c]{ll}%
\left[  J_{ab},J_{cd}\right]  =f_{ab,cd}^{\text{ \ \ \ \ \ \ \ }ef}J_{ef}, &
\left[  Z_{ab},P_{c}\right]  =f_{ab,c}^{\text{ \ \ \ \ }d}Z_{d},\\
\left[  J_{ab},Z_{cd}\right]  =f_{ab,cd}^{\text{ \ \ \ \ \ \ \ }ef}Z_{ef}, &
\left[  Z_{ab},Z_{c}\right]  =-f_{ab,c}^{\text{ \ \ \ \ }d}P_{d},\\
\left[  Z_{ab},Z_{cd}\right]  =-f_{ab,cd}^{\text{ \ \ \ \ \ \ }ef}J_{ef}, &
\left[  P_{a},P_{b}\right]  =J_{ab},\\
\left[  J_{ab},P_{c}\right]  =f_{ab,c}^{\text{ \ \ \ \ \ }d}P_{d}, & \left[
P_{a},Z_{b}\right]  =Z_{ab},\\
\left[  J_{ab},Z_{c}\right]  =f_{ab,c}^{\text{ \ \ \ \ }d}Z_{d}, & \left[
Z_{a},Z_{b}\right]  =-J_{ab},
\end{array}
\label{conmc}%
\end{equation}
where we can see the the $AdS_{5}$ algebra is a subalgebra of the
$\mathfrak{C}_{5}$-algebra, i.e.,
\begin{equation}
AdS_{5}\subset\mathfrak{C}_{5}.\label{31}%
\end{equation}

It is interesting to note that the commutation relations of the algebra
$Z_{2}\times AdS_{5}$ are similar to the commutation relations of the algebra
$\mathfrak{C}_{5}$. The generators of the algebra $Z_{2}\times AdS_{5}$
satisfy the following commutation relations:
\[%
\begin{array}
[c]{ll}%
\left[  J_{ab},J_{cd}\right]  =f_{ab,cd}^{\text{ \ \ \ \ \ \ }ef}J_{ef}, &
\left[  Z_{ab},P_{c}\right]  =f_{ab,c}^{\text{ \ \ \ \ }d}Z_{d},\\
\left[  J_{ab},Z_{cd}\right]  =f_{ab,cd}^{\text{ \ \ \ \ \ \ }ef}Z_{ef}, &
\left[  Z_{ab},Z_{c}\right]  =f_{ab,c}^{\text{ \ \ \ \ }d}P_{d},\\
\left[  Z_{ab},Z_{cd}\right]  =f_{ab,cd}^{\text{ \ \ \ \ \ \ }ef}J_{ef}, &
\left[  P_{a},P_{b}\right]  =J_{ab},\\
\left[  J_{ab},P_{c}\right]  =f_{ab,c}^{\text{ \ \ \ \ }d}P_{d}, & \left[
P_{a},Z_{b}\right]  =Z_{ab},\\
\left[  J_{ab},Z_{c}\right]  =f_{ab,c}^{\text{ \ \ \ }d}Z_{d}, & \left[
Z_{a},Z_{b}\right]  =J_{ab}.
\end{array}
\]
The natural question is: what is the advantage of using the algebra
$\mathfrak{C}_{5}\simeq\left(  Z_{4}\times AdS_{5}\right)  _{H}$ instead of
$Z_{2}\times AdS_{5}$?. The difference between the two algebras is in
invariant tensors. Indeed, the only invariant tensor algebra $AdS_{5}$ is
given by:
\[
\left\langle \tilde{J}_{ab},\tilde{J}_{cd},\tilde{P}_{e}\right\rangle
=\frac{1}{8}\varepsilon_{abcde},
\]
Using the theorem 2 (see Section 5) we find that the tensor invariants for
the algebra $\mathfrak{C}_{5}\simeq\left(  Z_{4}\times AdS_{5}\right)  _{H}$
are given by
\begin{align}
\left\langle J_{(ab,i)},J_{(cd,j)},P_{(e,k)}\right\rangle  &  =\alpha_{\gamma
}K_{ijk}^{\gamma}\left\langle \tilde{J}_{ab},\tilde{J}_{cd},\tilde{P}%
_{e}\right\rangle ,\nonumber\\
&  =\frac{1}{8}\left(  \alpha_{0}K_{ijk}^{0}+\alpha_{1}K_{ijk}^{1}+\alpha
_{2}K_{ijk}^{2}+\alpha_{3}K_{ijk}^{3}\right)  \varepsilon_{abcde}.\label{32}%
\end{align}
From (\ref{eq:z4}) can see that the only $3$-selectors nonzero, for
$i,j,k=0,1$, are $K_{000}^{0}$, $K_{100}^{1}$, $K_{010}^{1}$, $K_{001}^{1}$,
$K_{110}{}^{2}$, $K_{011}{}^{2}$, $K_{101}{}^{2}`$ y $K_{111}{}^{3}$. \ 

On the other hand, the invariant tensors for algebra $Z_{2}\times AdS_{5}$ are
given by (see theorem 2, section 5)
\begin{align}
\left\langle J_{(ab,\alpha)},J_{(cd,\beta)},P_{(e,\gamma)}\right\rangle  &
=\alpha_{\delta}K_{\alpha\beta\gamma}^{\delta}\left\langle \tilde{J}%
_{ab},\tilde{J}_{cd},\tilde{P}_{e}\right\rangle ,\nonumber\\
&  =\frac{1}{8}\left(  \alpha_{0}K_{\alpha\beta\gamma}^{0}+\alpha_{1}%
K_{\alpha\beta\gamma}^{1}\right)  \varepsilon_{abcde}. \label{33}%
\end{align}
From the law of multiplication of $Z_{2}$ we can see that the only
$3$-selectors nonzero for $\alpha,\beta,\gamma=0,1,$ are $K_{000}^{0}$,
$K_{100}^{1}$, $K_{010}^{1}$, $K_{001}^{1}$, $K_{110}{}^{0}$, $K_{011}{}^{0}$,
$K_{101}{}^{0}$ and $K_{111}{}^{1}$.

From (\ref{32}) we see that the choice $\alpha_{2}=\alpha_{3}=0$ leads to the
invariant tensors for algebra $\mathfrak{B}_{5}$ of \cite{standardg}, namely
$\left\langle J_{ab},J_{cd},P_{e}\right\rangle $, $\left\langle Z_{ab}%
,J_{cd},P_{e}\right\rangle $, $\left\langle J_{ab},Z_{cd},P_{e}\right\rangle
$, $\left\langle J_{ab},J_{cd},Z_{e}\right\rangle $. From (\ref{32}) we see
that the choice $\alpha_{1}=0$ leads to the following invariant tensors
$\left\langle J_{ab},J_{cd},P_{e}\right\rangle $, $\left\langle Z_{ab}%
,Z_{cd},P_{e}\right\rangle $, $\left\langle J_{ab},Z_{cd},Z_{e}\right\rangle
$, $\left\langle Z_{ab},J_{cd},Z_{e}\right\rangle $.

It is of interest to note that the existence of four arbitrary constants in
(\ref{32}) allows us to choose invariant tensors that are not possible for
$Z_{2}\times\mathfrak{g}$. For example the choice of the constants
\begin{equation}
\left(  \alpha_{0},\alpha_{1},\alpha_{2},\alpha_{3}\right)  =\alpha_{0}\left(
1,-1,-1,-1\right)  ,
\end{equation}
leads to interesting Chern-Simons Lagrangian for gravitation.

From (\ref{32}) we see that nonzero tensor invariants for the algebra
$\mathfrak{C}_{5}$ are given by (the factor 1/8 is abosorbed in $\alpha$)
\begin{equation}%
\begin{array}
[c]{l}%
\left\langle J_{ab},J_{cd},P_{e}\right\rangle =\alpha_{0}\,\varepsilon
_{abcde},\\
\left\langle J_{ab},J_{cd},Z_{e}\right\rangle =\alpha_{1}\,\varepsilon
_{abcde},\\
\left\langle J_{ab},Z_{cd},P_{e}\right\rangle =\alpha_{1}\,\varepsilon
_{abcde},\\
\left\langle J_{ab},Z_{cd},Z_{e}\right\rangle =\alpha_{2}\,\varepsilon
_{abcde},\\
\left\langle Z_{ab},Z_{cd},P_{e}\right\rangle =\alpha_{2}\,\varepsilon
_{abcde},\\
\left\langle Z_{ab},Z_{cd},Z_{e}\right\rangle =\alpha_{3}\,\varepsilon
_{abcde}.
\end{array}
\label{34}%
\end{equation}

For simplicity it is convenient to perform the following change of basis
\begin{align}
\bar{P}_{a}  &  =\frac{1}{\sqrt{2}}P_{a}+\frac{1}{\sqrt{2}}Z_{a},\nonumber\\
\bar{Z}_{a}  &  =\frac{1}{\sqrt{2}}P_{a}-\frac{1}{\sqrt{2}}Z_{a}. \label{35}%
\end{align}
In this basis the commutation relations take the form
\begin{align*}
\left[  J_{ab},\bar{P}_{c}\right]   &  =\frac{1}{\sqrt{2}}\left[  J_{ab}%
,P_{c}\right]  +\frac{1}{\sqrt{2}}\left[  J_{ab},Z_{c}\right]  ,\\
&  =\frac{1}{\sqrt{2}}f_{ab,c}^{d}P_{d}+\frac{1}{\sqrt{2}}f_{ab,c}^{d}Z_{d},\\
&  =f_{ab,c}^{d}\bar{P}_{d},
\end{align*}%
\begin{align*}
\left[  J_{ab},\bar{Z}_{c}\right]   &  =\frac{1}{\sqrt{2}}\left[  J_{ab}%
,P_{c}\right]  -\frac{1}{\sqrt{2}}\left[  J_{ab},Z_{c}\right]  ,\\
&  =\frac{1}{\sqrt{2}}f_{ab,c}^{d}P_{d}-\frac{1}{\sqrt{2}}f_{ab,c}^{d}Z_{d},\\
&  =f_{ab,c}^{d}\bar{Z}_{d},
\end{align*}%
\begin{align*}
\left[  Z_{ab},\bar{P}_{c}\right]   &  =\frac{1}{\sqrt{2}}\left[  Z_{ab}%
,P_{c}\right]  +\frac{1}{\sqrt{2}}\left[  Z_{ab},Z_{c}\right]  ,\\
&  =\frac{1}{\sqrt{2}}f_{ab,c}^{d}Z_{d}-\frac{1}{\sqrt{2}}f_{ab,c}^{d}P_{d},\\
&  =-f_{ab,c}^{d}\bar{Z}_{d},
\end{align*}%
\begin{align*}
\left[  Z_{ab},\bar{Z}_{c}\right]   &  =\frac{1}{\sqrt{2}}\left[  Z_{ab}%
,P_{c}\right]  -\frac{1}{\sqrt{2}}\left[  Z_{ab},Z_{c}\right]  ,\\
&  =\frac{1}{\sqrt{2}}f_{ab,c}^{d}Z_{d}+\frac{1}{\sqrt{2}}f_{ab,c}^{d}P_{d},\\
&  =f_{ab,c}^{d}\bar{P}_{d},
\end{align*}%
\begin{align*}
\left[  \bar{P}_{a},\bar{P}_{b}\right]   &  =\frac{1}{2}\left[  P_{a}%
+Z_{a},P_{b}+Z_{b}\right]  ,\\
&  =\frac{1}{2}\left(  \left[  P_{a},P_{b}\right]  +\left[  P_{a}%
,Z_{b}\right]  +\left[  Z_{a},P_{b}\right]  +\left[  Z_{a},Z_{b}\right]
\right)  ,\\
&  =\frac{1}{2}\left(  J_{ab}+Z_{ab}-Z_{ba}-J_{ab}\right)  ,\\
&  =Z_{ab},
\end{align*}%
\begin{align*}
\left[  \bar{P}_{a},\bar{Z}_{b}\right]   &  =\frac{1}{2}\left[  P_{a}%
+Z_{a},P_{b}-Z_{b}\right]  ,\\
&  =\frac{1}{2}\left(  \left[  P_{a},P_{b}\right]  -\left[  P_{a}%
,Z_{b}\right]  +\left[  Z_{a},P_{b}\right]  -\left[  Z_{a},Z_{b}\right]
\right)  ,\\
&  =\frac{1}{2}\left(  J_{ab}-Z_{ab}-Z_{ba}+J_{ab}\right)  ,\\
&  =J_{ab},
\end{align*}%
\begin{align*}
\left[  \bar{Z}_{a},\bar{Z}_{b}\right]   &  =\frac{1}{2}\left[  P_{a}%
-Z_{a},P_{b}-Z_{b}\right]  ,\\
&  =\frac{1}{2}\left(  \left[  P_{a},P_{b}\right]  -\left[  P_{a}%
,Z_{b}\right]  -\left[  Z_{a},P_{b}\right]  +\left[  Z_{a},Z_{b}\right]
\right)  ,\\
&  =\frac{1}{2}\left(  J_{ab}-Z_{ab}+Z_{ba}-J_{ab}\right)  ,\\
&  =-Z_{ab}.
\end{align*}
Moreover, the invariant tensors takes the form (the factor $1/\sqrt{2}$ is
absorbed in the coeficients $\alpha$)
\begin{align*}
\left\langle J_{ab},J_{cd},\bar{P}_{e}\right\rangle  &  =\frac{1}{\sqrt{2}%
}\left\langle J_{ab},J_{cd},P_{e}\right\rangle +\frac{1}{\sqrt{2}}\left\langle
J_{ab},J_{cd},Z_{e}\right\rangle ,\\
&  =\left(  \alpha_{0}+\alpha_{1}\right)  \varepsilon_{abcde},
\end{align*}%
\begin{align*}
\left\langle J_{ab},J_{cd},\bar{Z}_{e}\right\rangle  &  =\frac{1}{\sqrt{2}%
}\left\langle J_{ab},J_{cd},P_{e}\right\rangle -\frac{1}{\sqrt{2}}\left\langle
J_{ab},J_{cd},Z_{e}\right\rangle ,\\
&  =\left(  \alpha_{0}-\alpha_{1}\right)  \varepsilon_{abcde},
\end{align*}%
\begin{align*}
\left\langle J_{ab},Z_{cd},\bar{P}_{e}\right\rangle  &  =\frac{1}{\sqrt{2}%
}\left\langle J_{ab},Z_{cd},P_{e}\right\rangle +\frac{1}{\sqrt{2}}\left\langle
J_{ab},Z_{cd},Z_{e}\right\rangle ,\\
&  =\left(  \alpha_{1}+\alpha_{2}\right)  \,\varepsilon_{abcde},
\end{align*}%
\begin{align*}
\left\langle J_{ab},Z_{cd},\bar{Z}_{e}\right\rangle  &  =\frac{1}{\sqrt{2}%
}\left\langle J_{ab},Z_{cd},P_{e}\right\rangle -\frac{1}{\sqrt{2}}\left\langle
J_{ab},Z_{cd},Z_{e}\right\rangle ,\\
&  =\left(  \alpha_{1}-\alpha_{2}\right)  \,\varepsilon_{abcde},
\end{align*}%
\begin{align*}
\left\langle Z_{ab},Z_{cd},\bar{P}_{e}\right\rangle  &  =\frac{1}{\sqrt{2}%
}\left\langle Z_{ab},Z_{cd},P_{e}\right\rangle +\frac{1}{\sqrt{2}}\left\langle
Z_{ab},Z_{cd},Z_{e}\right\rangle ,\\
&  =\left(  \alpha_{2}+\alpha_{3}\right)  \,\varepsilon_{abcde},
\end{align*}%
\begin{align*}
\left\langle Z_{ab},Z_{cd},\bar{Z}_{e}\right\rangle  &  =\frac{1}{\sqrt{2}%
}\left\langle Z_{ab},Z_{cd},P_{e}\right\rangle -\frac{1}{\sqrt{2}}\left\langle
Z_{ab},Z_{cd},Z_{e}\right\rangle ,\\
&  =\left(  \alpha_{2}-\alpha_{3}\right)  \,\varepsilon_{abcde},
\end{align*}

In summary, in the bases (\ref{35}) the commutation relations and the
invariant tensor for algebra $\mathfrak{C}_{5}$ are given by
\begin{equation}%
\begin{array}
[c]{ll}%
\left[  J_{ab},J_{cd}\right]  =f_{ab,cd}^{ef}J_{ef}, & \left[  Z_{ab}%
,P_{c}\right]  =-f_{ab,c}^{d}Z_{d},\\
\left[  J_{ab},Z_{cd}\right]  =f_{ab,cd}^{ef}Z_{ef}, & \left[  Z_{ab}%
,Z_{c}\right]  =f_{ab,c}^{d}P_{d},\\
\left[  Z_{ab},Z_{cd}\right]  =-f_{ab,cd}^{ef}J_{ef}, & \left[  P_{a}%
,P_{b}\right]  =Z_{ab},\\
\left[  J_{ab},P_{c}\right]  =f_{ab,c}^{d}P_{d}, & \left[  P_{a},Z_{b}\right]
=J_{ab},\\
\left[  J_{ab},Z_{c}\right]  =f_{ab,c}^{d}Z_{d}, & \left[  Z_{a},Z_{b}\right]
=-Z_{ab},
\end{array}
\label{eq:relconmcesc}%
\end{equation}%
\begin{equation}%
\begin{array}
[c]{l}%
\left\langle J_{ab},J_{cd},P_{e}\right\rangle =\left(  \alpha_{0}+\alpha
_{1}\right)  \,\varepsilon_{abcde},\\
\left\langle J_{ab},J_{cd},Z_{e}\right\rangle =\left(  \alpha_{0}-\alpha
_{1}\right)  \,\varepsilon_{abcde},\\
\left\langle J_{ab},Z_{cd},P_{e}\right\rangle =\left(  \alpha_{1}+\alpha
_{2}\right)  \,\varepsilon_{abcde},\\
\left\langle J_{ab},Z_{cd},Z_{e}\right\rangle =\left(  \alpha_{1}-\alpha
_{2}\right)  \,\varepsilon_{abcde},\\
\left\langle Z_{ab},Z_{cd},P_{e}\right\rangle =\left(  \alpha_{2}+\alpha
_{3}\right)  \,\varepsilon_{abcde},\\
\left\langle Z_{ab},Z_{cd},Z_{e}\right\rangle =\left(  \alpha_{2}-\alpha
_{3}\right)  \,\varepsilon_{abcde}.
\end{array}
\label{eq:tensinvcesc}%
\end{equation}
where we made {}{}the change of rotulo $\bar{P}_{a}\rightarrow P_{a}$,
$\bar{Z}_{a}\rightarrow Z_{a}$

\section{\textbf{The Chern-Simons Lagrangian invariant under }$\mathfrak{C}%
_{5}$}

Using the subspace separation method introduced in Ref. \cite{sepsubesp} it is
possible to find the Chern-Simons Lagrangian in five dimensions for the
$\mathfrak{C}_{5}$ algebra. \ Since this algebra is generated by
$J_{ab},Z_{ab},P_{a},Z_{a}$ \ we can identify the relevant subspaces present
in the $\mathfrak{C}_{5}$ algebra, i.e.,
\[
\mathfrak{C}_{5}=\left\{  J_{ab}\right\}  \oplus\left\{  Z_{ab}\right\}
\oplus\left\{  P_{a}\right\}  \oplus\left\{  Z_{a}\right\}  ,
\]
and write the connections in terms of pieces valued on every subspace, i.e.,%

\begin{equation}%
\begin{array}
[c]{l}%
A=\omega+e+k+h,\\
A_{2}=\omega+e,\\
A_{1}=\omega,\\
\bar{A}=0,
\end{array}
\label{eq:conc}%
\end{equation}

Since $Q^{(5)}(A,\bar{A}=0)=Q^{(5)}(A)$, the Chern-Simons Lagrangian is given
by
\begin{equation}
\mathcal{L}_{\text{CS},\mathfrak{C}_{5}}(A)=\kappa\,Q^{(5)}(A)=\kappa
\,Q^{(5)}(A,\bar{A}=0) \label{36}%
\end{equation}
where
\[
\omega=\frac{1}{2}\omega^{ab}J_{ab},\quad k=\frac{1}{2}k^{ab}Z_{ab},\quad
e=\frac{1}{\ell}e^{a}P_{a},\quad h=\frac{1}{\ell}h^{a}Z_{a}.
\]

Repeated use of the triangle equation \cite{sepsubesp} allows us to split the
Lagrangian as
\begin{equation}
Q^{(5)}(A,\bar{A})=Q^{(5)}(A,A_{2})+Q^{(5)}(A_{2},A_{1})+Q^{(5)}(A_{1},\bar
{A})+dB, \label{37}%
\end{equation}
where $B$ is a $4$-form given by%
\begin{equation}
B=Q^{(4)}(A_{2},A_{1},\bar{A})+Q^{(4)}(A,A_{2},\bar{A}). \label{38}%
\end{equation}
which correspond to a boundary term.

The calculation of the transgression forms, $Q^{(5)}(A,A_{2})$, $Q^{(5)}%
(A_{2},A_{1})$ and $Q^{(5)}(A_{1},\bar{A})$, is carried out in detail in the appendix.

The choice of arbitrary coefficients $\alpha$ in the form (see Appendix 2, eq.
(\ref{a1}))
\begin{equation}
\left(  \alpha_{0},\alpha_{1},\alpha_{2},\alpha_{3}\right)  =\alpha_{0}\left(
1,-1,-1,-1\right)  . \label{39}%
\end{equation}
leads (modulo boundary terms) the following Lagrangian
\begin{align}
\mathcal{L}_{\text{ChS}_{5},\mathfrak{C}_{5}}  &  =\alpha_{0}\,\varepsilon
_{abcde}\left(  R^{ab}e^{c}e^{d}e^{e}+\frac{3}{10\ell^{2}}e^{a}e^{b}e^{c}%
e^{d}e^{e}-\frac{3}{2}\ell^{2}k^{ab}R^{cd}T^{e}-\frac{1}{2}\ell^{2}%
R^{ab}k^{cd}k_{f}^{e}h^{f}\right. \nonumber\\
&  -\frac{3}{2}k^{ab}T^{c}e^{d}e^{e}-\frac{3}{4}\ell^{2}k^{ab}T^{c}%
\mathrm{D}_{\omega}k^{de}-\frac{3}{2}k^{ab}T^{c}e^{d}h^{e}+\frac{1}{2}\ell
^{2}k^{ab}T^{c}`k_{f}^{d}`k^{fe}\nonumber\\
&  +\frac{1}{2}k^{ab}T^{c}h^{d}h^{e}-\frac{1}{2}k^{ab}e^{c}e^{d}k_{f}^{e}%
h^{f}-\frac{3}{8}\ell^{2}k^{ab}\mathrm{D}_{\omega}k^{cd}k_{f}^{e}`h^{f}%
-\frac{3}{4}k^{ab}e^{c}h^{d}k_{f}^{e}h^{f}\nonumber\\
&  +\frac{3}{10}\ell^{2}k^{ab}k_{f}^{c}k^{fd}k_{g}^{e}h^{g}+\frac{3}{10}%
k^{ab}h^{c}h^{d}k_{f}^{e}k^{f}+\frac{3}{4}\ell^{2}R^{ab}R^{cd}h^{e}+\frac
{1}{\ell^{2}}e^{a}h^{b}e^{c}h^{d}h^{e}\nonumber\\
&  \left.  +\frac{3}{20}\ell^{2}k_{f}^{a}k^{fb}k_{g}^{c}k^{gd}h^{e}+\frac
{3}{2}R^{ab}e^{c}h^{d}h^{e}-\frac{3}{4}k_{f}^{a}k^{fb}e^{c}h^{d}h^{e}-\frac
{1}{2}\ell^{2}R^{ab}k_{f}^{c}k^{fd}h^{e}\right)  , \label{40}%
\end{align}
where the term $-\kappa/\ell^{3}$ is reabsorbed in the coefficient $\alpha
_{0}$.

From (\ref{40}) we can see that:

$\left(  i\right)  $ $\ \mathcal{L}_{\text{ChS}_{5},\mathfrak{C}_{5}}$
contains the Einstein-Hilbert Lagrangian with a cosmological terms

$\left(  ii\right)  $ for $k^{ab}=0$ we have
\begin{equation}
\mathcal{L}_{\text{ChS}_{5},\mathfrak{C}_{5}}=\alpha_{0}\,\varepsilon
_{abcde}\left(  R^{ab}e^{c}e^{d}e^{e}+\frac{3}{10\ell^{2}}e^{a}e^{b}e^{c}%
e^{d}e^{e}+\frac{3}{4}\ell^{2}R^{ab}R^{cd}h^{e}+\frac{1}{\ell^{2}}e^{a}%
h^{b}e^{c}h^{d}h^{e}+\frac{3}{2}R^{ab}e^{c}h^{d}h^{e}\right)  ,
\end{equation}
from where we can see that, in the particular case that $e^{a}=h^{a}$ ,
$\mathcal{L}_{\text{ChS}_{5},\mathfrak{C}_{5}}$ takes the form
\[
\mathcal{L}_{\text{ChS}_{5},\mathfrak{C}_{5}}=\frac{\alpha_{0}}{2}%
\,\varepsilon_{abcde}\left(  5R^{ab}e^{c}e^{d}e^{e}+\frac{13}{5\ell^{2}}%
e^{a}e^{b}e^{c}e^{d}e^{e}+\frac{3}{2}\ell^{2}R^{ab}R^{cd}e^{e}\right)
\]
that in the case that $\ell<<1$ we have the Einstein-Hilbert Lagrangian with
cosmological term.

$\left(  iii\right)  $ for $h^{a}=0$ we find
\begin{align}
\mathcal{L}_{\text{CS}_{5},\mathfrak{C}_{5}}  &  =\alpha_{0}\,\varepsilon
_{abcde}\left(  R^{ab}e^{c}e^{d}e^{e}+\frac{3}{10\ell^{2}}e^{a}e^{b}e^{c}%
e^{d}e^{e}-\frac{3}{2}\ell^{2}k^{ab}R^{cd}T^{e}-\frac{3}{2}k^{ab}T^{c}%
e^{d}e^{e}-\frac{3}{4}\ell^{2}k^{ab}T^{c}\mathrm{D}_{\omega}k^{de}\right.
\nonumber\label{eq:h0}\\
&  +\left.  \frac{1}{2}\ell^{2}k^{ab}T^{c}k_{f}^{d}k^{fe}\right)  .
\end{align}
from where we see that if we impose the torsion free condition, then we obtain
the Einstein-Hilbert Lagrangian with cosmological term.

$\left(  iv\right)  $ When $k^{ab}=0$ and $h^{a}=0$ we have the Einstein-Hilbert Lagrangian with a cosmological term.

\section{\textbf{ (2 +1)-dimensional case}}

The $\mathfrak{C}_{5}=\left(  Z_{4}\times AdS_{5}\right)  _{H}$ algebra can be
generalized to the case of arbitrary dimensions, i.e., it can be generalized
to the case%

\begin{equation}
\mathfrak{C}_{d}=\left(  Z_{4}\times\mathfrak{ads}_{d}\right)  _{H},
\label{eq:algebracd}%
\end{equation}
whose generators satisfy the following commutation relations
\begin{equation}%
\begin{array}
[c]{ll}%
\left[  J_{ab},J_{cd}\right]  =f_{ab,cd}^{ef}J_{ef}, & \left[  Z_{ab}%
,P_{c}\right]  =f_{ab,c}^{d}Z_{d},\\
\left[  J_{ab},Z_{cd}\right]  =f_{ab,cd}^{ef}Z_{ef}, & \left[  Z_{ab}%
,Z_{c}\right]  =-f_{ab,c}^{d}P_{d},\\
\left[  Z_{ab},Z_{cd}\right]  =-f_{ab,cd}^{ef}`J_{ef}, & \left[  P_{a}%
,P_{b}\right]  =J_{ab},\\
\left[  J_{ab},P_{c}\right]  =f_{ab,c}^{d}P_{d}, & \left[  P_{a},Z_{b}\right]
=Z_{ab},\\
\left[  J_{ab},Z_{c}\right]  =f_{ab,c}^{d}Z_{d}, & \left[  Z_{a},Z_{b}\right]
=-J_{ab}.
\end{array}
\label{conmcd}%
\end{equation}

The only nonzero invariant tensor for the $AdS_{3}=\mathfrak{so}(2,2)$ algebra
is given by
\[
\left\langle J_{ab}P_{c}\right\rangle =\frac{1}{4}\varepsilon_{abc}%
\]

Using the theorem of Section 5 is found that
\begin{equation}
\left\langle J_{(ab,i)},P_{(c,j)}\right\rangle =\frac{1}{2}\left(  \alpha
_{0}K_{ij}{}^{0}+\alpha_{1}K_{ij}{}^{1}+\alpha_{2}K_{ij}{}^{2}+\alpha
_{3}K_{ij}{}^{3}\right)  \varepsilon_{abc},
\end{equation}
are the invariant tensors for algebra $\mathfrak{C}_{3}$. \ For $Z_{4}$ group,
the only two-selectors non-zero , with $i,j=0,1$ are $K_{00}{}^{0}$, $K_{01}%
{}^{1}$, $K_{10}{}^{1}$ y $K_{11}{}^{2}$. So that the corresponding invariant
tensors are given by
\begin{equation}%
\begin{array}
[c]{l}%
\left\langle J_{ab},P_{c}\right\rangle =\alpha_{0}\,\varepsilon_{abc},\\
\left\langle J_{ab},Z_{c}\right\rangle =\alpha_{1}\,\varepsilon_{abc},\\
\left\langle Z_{ab},P_{c}\right\rangle =\alpha_{1}\,\varepsilon_{abc},\\
\left\langle Z_{ab},Z_{c}\right\rangle =\alpha_{2}\,\varepsilon_{abc}.
\end{array}
\label{eq:tensinvc3}%
\end{equation}

Under the change of basis
\begin{align}
P_{a}  &  \rightarrow\frac{1}{\sqrt{2}}P_{a}+\frac{1}{\sqrt{2}}Z_{a}%
,\label{eq:cdbd2}\\
Z_{a}  &  \rightarrow\frac{1}{\sqrt{2}}P_{a}-\frac{1}{\sqrt{2}}Z_{a},
\end{align}
the commutation relations take the form
\begin{equation}%
\begin{array}
[c]{ll}%
\left[  J_{ab},J_{cd}\right]  =f_{ab,cd}^{ef}J_{ef}, & \left[  Z_{ab}%
,P_{c}\right]  =-f_{ab,c}^{d}Z_{d},\\
\left[  J_{ab},Z_{cd}\right]  =f_{ab,cd}^{ef}Z_{ef}, & \left[  Z_{ab}%
,Z_{c}\right]  =f_{ab,c}^{d}P_{d},\\
\left[  Z_{ab},Z_{cd}\right]  =-f_{ab,cd}^{ef}J_{ef}, & \left[  P_{a}%
,P_{b}\right]  =Z_{ab},\\
\left[  J_{ab},P_{c}\right]  =f_{ab,c}^{d}P_{d}, & \left[  P_{a},Z_{b}\right]
=J_{ab},\\
\left[  J_{ab},Z_{c}\right]  =f_{ab,c}^{d}Z_{d}, & \left[  Z_{a},Z_{b}\right]
=-Z_{ab}.
\end{array}
\label{eq:relconmc3esc}%
\end{equation}
and corresponding invariant tensors are:
\begin{equation}%
\begin{array}
[c]{l}%
\left\langle J_{ab},P_{c}\right\rangle =\left(  \alpha_{0}+\alpha_{1}\right)
\,\varepsilon_{abc},\\
\left\langle J_{ab},Z_{c}\right\rangle =\left(  \alpha_{0}-\alpha_{1}\right)
\,\varepsilon_{abc},\\
\left\langle Z_{ab},P_{c}\right\rangle =\left(  \alpha_{1}+\alpha_{2}\right)
\,\varepsilon_{abc},\\
\left\langle Z_{ab},Z_{c}\right\rangle =\left(  \alpha_{1}-\alpha_{2}\right)
\,\varepsilon_{abc}.
\end{array}
\label{eq:tensinvc3esc}%
\end{equation}

For the Chern-Simons Lagrangian
\begin{equation}
\mathcal{L}_{\text{CS}_{3},\mathfrak{c_{3}}}=\kappa\,Q^{(3)}(A,0)=\kappa
\,Q^{(3)}(A),
\end{equation}
we use the method of separation of subspaces. Following the same procedure
used in case $5$-dimensional we find
\begin{align}
\mathcal{L}_{\text{CS}_{3},\mathfrak{c_{3}}}  &  =\varepsilon_{abc}\left(
\left(  \alpha_{0}+\alpha_{1}\right)  R^{ab}e^{c}+\frac{1}{3\ell^{2}}\left(
\alpha_{1}+\alpha_{2}\right)  e^{a}e^{b}e^{c}+\left(  \alpha_{1}+\alpha
_{2}\right)  k^{ab}T^{c}\right. \nonumber\\
&  +\left(  \alpha_{1}-\alpha_{2}\right)  \frac{1}{2}k^{ab}\mathrm{D}_{\omega
}h^{c}-\frac{1}{2}\left(  \alpha_{1}-\alpha_{2}\right)  k^{ab}k_{d}^{c}%
e^{d}+\frac{1}{3}\left(  \alpha_{1}+\alpha_{2}\right)  k^{ab}k_{d}^{c}%
h^{d}\nonumber\\
&  +\left(  \alpha_{0}-\alpha_{1}\right)  R^{ab}h^{c}+\frac{1}{\ell^{2}%
}\left(  \alpha_{1}-\alpha_{2}\right)  e^{a}e^{b}h^{c}+\frac{1}{2}\left(
\alpha_{1}-\alpha_{2}\right)  \mathrm{D}_{\omega}k^{ab}h^{c}\nonumber\\
&  +\left.  \frac{1}{\ell^{2}}\left(  \alpha_{0}-\alpha_{1}\right)  e^{a}%
h^{b}h^{c}-\frac{1}{3}\left(  \alpha_{0}-\alpha_{1}\right)  k_{d}^{a}%
k^{db}h^{c}-\frac{1}{3\ell^{2}}\left(  \alpha_{1}-\alpha_{2}\right)
h^{a}h^{b}h^{c}\right)  . \label{41}%
\end{align}

From (\ref{41}) we can see that:

$\left(  i\right)  $ \ When $k^{ab}=0$ and $h^{a}=0$ we have the
Einstein-Hilbert Lagrangian with cosmological term for certain values {}{}of
the constant $\alpha$.

The arbitrariness of the constant $\alpha$ in terms of Einstein-Hilbert and
the cosmological constant allow interpreting the Lagrangian as the
Lanczos-Lovelock Lagrangian in $3$ dimensions

This occurs even in cases where $\alpha_{0}-\alpha_{1}=0$, $\alpha_{1}=0$ y
$\alpha_{1}-\alpha_{2}=0$. This result allows us to interpret the Lagrangian
(\ref{41}) as a Lagrangian describing a link between Lovelock gravity and
bosonic fields $k^{ab}$ and $h^{a}$.

$\left(  ii\right)  $ In the case that $h^{a}=0$ we have
\begin{align}
&  \mathcal{L}_{\text{CS}_{3},C\mathfrak{_{3}}}\nonumber\\
&  =\varepsilon_{abc}\left(  \left(  \alpha_{0}+\alpha_{1}\right)  R^{ab}%
e^{c}+\frac{1}{3\ell^{2}}\left(  \alpha_{1}+\alpha_{2}\right)  e^{a}e^{b}%
e^{c}+\left(  \alpha_{1}+\alpha_{2}\right)  k^{ab}T^{c}-\frac{1}{2}\left(
\alpha_{1}-\alpha_{2}\right)  k^{ab}k_{d}^{c}e^{d}\right)  .
\end{align}
Note that if $T^{a}=0$ and $\alpha_{1}=\alpha_{2}$, then we obtain the
Einstein-Hilbert Lagrangian with cosmological term.

\section{\textbf{Comments}}

In the present article we have introduced a modification to the Lie algebra
$S$-expansion procedure. \ The modification is carried out by imposing a
condition, which was called $H$-condition, on the $S$-expansion procedure,
when the semigroup is given by a cyclic group of even order.\ The invariant
tensors for $S_{H}$-expanded algebras are calculated and the\ dual formulation
of $S_{H}$-expansion procedure was found. \ 

Following Ref. \cite{standardg}, we\ have considered the $S_{H}$-expansion of
the five-dimensional $AdS$ algebra and its corresponding invariants tensors
were obtained. \ Then a Chern-Simons Lagrangian invariant under the
five-dimensional $AdS$ algebra $S_{H}$-expanded is constructed and its
relationship to the general relativity was studied.

This work was supported in part by $FONDECYT$ through Grant N$^{0}$ 1130653.
Three of the authors ($NG$, $GR$, $SS$) were supported by grants from the
Comisi\'{o}n Nacional de Investigaci\'{o}n Cient\'{\i}fica y Tecnol\'{o}gica
CONICYT and from the Universidad de Concepci\'{o}n, Chile.

\section{\textbf{Appendix 1: \ Cyclic groups}}

\subsection{\textbf{Semigroups}}

A semigroup is a closed algebraic structure which also satisfies the axiom of
associativity:
\begin{equation}
\forall\,a,\,b,\,c\,\in A,\ a\ast(b\ast c)=(a\ast b)\ast c. \label{eq:axasoc}%
\end{equation}
For an associative structure no matter in what manner we perform the partial
products of an expression. Therefore it is possible to define a $r$-selector
to a semigroup of $\left\vert S\right\vert $ elements:
\begin{equation}
\lambda_{\alpha_{1}}\cdots\lambda_{\alpha_{r}}=\lambda_{\rho(\alpha_{1}%
,\ldots,\alpha_{r})}=K_{\alpha_{1}\ldots\alpha_{r}}^{\gamma}\lambda_{\gamma}.
\label{eq:rselector}%
\end{equation}
For an abelian semigroup, we have $\lambda_{\alpha}\lambda_{\beta}%
=\lambda_{\beta}\lambda_{\alpha}$ so that $\lambda_{\rho(\alpha,\beta
)}=\lambda_{\rho(\beta,\alpha)}$. This mean that
\begin{equation}
K_{\alpha\beta}{}^{\gamma}=K_{\beta\alpha}{}^{\gamma}. \label{eq:simselec}%
\end{equation}
In general $K_{\alpha_{1}\ldots\alpha_{r}}^{\gamma}$ is completely symmetric
in their lower indices. Consequently the useful identity is true:
\begin{equation}
K_{\alpha\beta\gamma}{}^{\delta}=K_{\beta\gamma\alpha}{}^{\delta}%
=K_{\gamma\alpha\beta}{}^{\delta}. \label{eq:simgselec}%
\end{equation}
Since
\begin{align*}
\lambda_{\alpha}\lambda_{\beta}  &  =K_{\alpha\beta}^{\gamma}\lambda_{\gamma
},\\
\ \quad\lambda_{\alpha}\lambda_{\beta}\lambda_{\delta}  &  =K_{\alpha\beta
}^{\gamma}\lambda_{\gamma}\lambda_{\delta},\\
K_{\alpha\beta\delta}^{\varepsilon}\lambda_{\varepsilon}  &  =K_{\alpha\beta
}^{\gamma}K_{\gamma\delta}^{\varepsilon}\lambda_{\varepsilon},
\end{align*}
we have
\begin{equation}
K_{\alpha\beta\gamma}{}^{\delta}=K_{\alpha\beta}{}^{\varepsilon}%
K_{\varepsilon\gamma}{}^{\delta} \label{eq:aumselec}%
\end{equation}
and in general
\begin{equation}
K_{\alpha_{1}\ldots\alpha_{r}}^{\gamma}=K_{\alpha_{1}\alpha_{2}}^{\beta_{1}%
}K_{\beta_{1}\alpha_{3}}^{\beta_{2}}\cdots K_{\beta_{r-3}\alpha_{r-1}}%
^{\beta_{r-2}}K_{\beta_{r-2}\alpha_{r}}^{\gamma}. \label{eq:aumselecg}%
\end{equation}

\subsection{\textbf{Direct product of semigroups}}

Let $S$ and $S^{\prime}$ be two semigroups. The Cartesian product $S\times
S^{\prime}$ together with the internal binary operation
\begin{equation}
(a,a^{\prime})(b,b^{\prime})=(ab,a^{\prime}b^{\prime}) \label{eq:prodsemi}%
\end{equation}
form a semigroup. Indeed, the closure property is satisfied by construction
(since $ab\in S$ and $a^{\prime}b^{\prime}\in S^{\prime}$). Let's check the
associative property:
\begin{align*}
(a,a^{\prime})[(b,b^{\prime})(c,c^{\prime})]  &  =(a,a^{\prime})(bc,b^{\prime
}c^{\prime}),\\
&  =\left(  a(bc),a^{\prime}(b^{\prime}c^{\prime})\right)  ,\\
&  =\left(  (ab)c,(a^{\prime}b^{\prime})c^{\prime}\right)  ,\\
&  =(ab,a^{\prime}b^{\prime})(c,c^{\prime}),\\
&  =[(a,a^{\prime})(b,b^{\prime})](c,c^{\prime}).
\end{align*}
This means that the associative property is satisfied, as a direct consequence
from the associativity of $S$ and $S^{\prime}$. In the representation of
selectors, we denote the internal law of the semigroup $S\times S^{\prime
}=\left\{  \lambda_{\alpha}\times\lambda_{\alpha^{\prime}}^{\prime}\right\}  $
as
\begin{equation}
\left(  \lambda_{\alpha}\times\lambda_{\alpha^{\prime}}^{\prime}\right)
\left(  \lambda_{\beta}\times\lambda_{\beta^{\prime}}^{\prime}\right)
=\left(  \lambda_{\alpha}\lambda_{\beta}\right)  \times\left(  \lambda
_{\alpha^{\prime}}^{\prime}\lambda_{\beta^{\prime}}^{\prime}\right)
=K_{\alpha\beta}^{\gamma}K_{\alpha^{\prime}\beta^{\prime}}^{\prime}{}%
^{\gamma^{\prime}}\lambda_{\gamma}\times\lambda_{\gamma^{\prime}}^{\prime}.
\label{eq:prodsemisel}%
\end{equation}

\subsection{\textbf{Cyclic groups}}

Cyclic groups $%
%TCIMACRO{\U{2124} }%
%BeginExpansion
\mathbb{Z}
%EndExpansion
_{n}$ are abelian groups whose elements can be expressed as a power of a
single element of the group. That is, if $a$ $\in$ $%
%TCIMACRO{\U{2124} }%
%BeginExpansion
\mathbb{Z}
%EndExpansion
_{n}$, then
\begin{equation}%
%TCIMACRO{\U{2124} }%
%BeginExpansion
\mathbb{Z}
%EndExpansion
_{n}=\left\{  a,aa,aaa,\ldots,a^{n}\right\}  .
\end{equation}
The group $%
%TCIMACRO{\U{2124} }%
%BeginExpansion
\mathbb{Z}
%EndExpansion
_{4}$ has the Cayley table

\[%
\begin{array}
[c]{|c||c|c|c|c|}\hline
Z_{4} & \overline{0} & \overline{1} & \overline{2} & \overline{3}%
\\\hline\hline
\overline{0} & \overline{0} & \overline{1} & \overline{2} & \overline
{3}\\\hline
\overline{1} & \overline{1} & \overline{2} & \overline{3} & \overline
{0}\\\hline
\overline{2} & \overline{2} & \overline{3} & \overline{0} & \overline
{1}\\\hline
\overline{3} & \overline{3} & \overline{0} & \overline{1} & \overline
{2}\\\hline
\end{array}
\]
or in the notation of selectors
\begin{equation}%
\begin{array}
[c]{|c||c|c|c|c|}\hline
Z_{4} & \lambda_{0} & \lambda_{1} & \lambda_{2} & \lambda_{3}\\\hline\hline
\lambda_{0} & \lambda_{0} & \lambda_{1} & \lambda_{2} & \lambda_{3}\\\hline
\lambda_{1} & \lambda_{1} & \lambda_{2} & \lambda_{3} & \lambda_{0}\\\hline
\lambda_{2} & \lambda_{2} & \lambda_{3} & \lambda_{0} & \lambda_{1}\\\hline
\lambda_{3} & \lambda_{3} & \lambda_{0} & \lambda_{1} & \lambda_{2}\\\hline
\end{array}
\label{eq:z4}%
\end{equation}

Some properties for cyclic groups of even order, $Z_{2n},$ are
\begin{align}
K_{k+n,l}{}^{m}  &  =\left\{
\begin{array}
[c]{ll}%
1, & m\equiv k+n+l\ \left(  \operatorname{mod}2n\right) \\
0, & otherwise
\end{array}
\right. \nonumber\label{eq:jacobi2}\\
&  =\left\{
\begin{array}
[c]{ll}%
1, & m+n\equiv k+{2n}+l\ \left(  \operatorname{mod}2n\right) \\
0, & otherwise
\end{array}
\right. \nonumber\\
&  =K_{kl}{}^{m+n}.
\end{align}%
\begin{align}
K_{k+n,l}{}^{m}  &  =\left\{
\begin{array}
[c]{ll}%
1, & m+n\equiv k+n+l\ \left(  \operatorname{mod}2n\right) \\
0, & otherwise
\end{array}
\right. \nonumber\label{eq:jacobi3}\\
&  =\left\{
\begin{array}
[c]{ll}%
1, & m+{2n}\equiv k+{2n}+l\ \left(  \operatorname{mod}2n\right) \\
0, & otherwise
\end{array}
\right. \nonumber\\
&  =K_{kl}{}^{m}.
\end{align}%
\begin{align}
K_{i+n,j+n}{}^{\gamma}  &  =\left\{
\begin{array}
[c]{ll}%
1, & \gamma\equiv i+n+j+n\ \left(  \operatorname{mod}2n\right) \\
0, & otherwise
\end{array}
\right. \nonumber\label{eq:fdual1}\\
&  =\left\{
\begin{array}
[c]{ll}%
1, & \gamma\equiv i+j+{2n}\ \left(  \operatorname{mod}2n\right) \\
0, & otherwise
\end{array}
\right. \nonumber\\
&  =K_{ij}{}^{\gamma}.
\end{align}%
\begin{align}
K_{i,j+n}{}^{k}  &  =\left\{
\begin{array}
[c]{ll}%
1, & k\equiv i+j+n\ \left(  \operatorname{mod}2n\right) \\
0, & otherwise
\end{array}
\right. \nonumber\label{eq:fdual2}\\
&  =\left\{
\begin{array}
[c]{ll}%
1, & k\equiv(i+n)+j\ \left(  \operatorname{mod}2n\right) \\
0, & otherwise
\end{array}
\right. \nonumber\\
&  =K_{i+n,j}{}^{k}.
\end{align}

\subsection{\textbf{The Klein group}}

It is possible to show that there are only two groups of four elements (up to
isomorphism): one is the cyclic group $Z_{4}$, and the other is the group of
Klein $D_{4}$. The Klein group $D_{4}$ corresponds to the direct product
$Z_{2}\times Z_{2}$. To prove this consider the multiplication rule of the
direct product of (semi) groups. If we denote the factor groups as $\left(
Z_{2},\bar{+}\right)  $ and as $\left(  Z_{2},\tilde{+}\right)  $,
\[%
\begin{array}
[c]{|c||c|c|}\hline
Z_{2} & \overline{0} & \overline{1}\\\hline\hline
\overline{0} & \overline{0} & \overline{1}\\\hline
\overline{1} & \overline{1} & \overline{0}\\\hline
\end{array}
\quad%
\begin{array}
[c]{|c||c|c|}\hline
Z_{2} & \tilde{0} & \tilde{1}\\\hline\hline
\tilde{0} & \tilde{0} & \tilde{1}\\\hline
\tilde{1} & \tilde{1} & \tilde{0}\\\hline
\end{array}
\]
we have
\[%
\begin{array}
[c]{|c||c|c|c|c|}\hline
Z_{2}\times Z_{2} & \left(  \overline{0},\tilde{0}\right)  & \left(
\overline{1},\tilde{0}\right)  & \left(  \overline{0},\tilde{1}\right)  &
\left(  \overline{1},\tilde{1}\right) \\\hline\hline
\left(  \overline{0},\tilde{0}\right)  & \left(  \overline{0},\tilde{0}\right)
& \left(  \overline{1},\tilde{0}\right)  & \left(  \overline{0},\tilde
{1}\right)  & \left(  \overline{1},\tilde{1}\right) \\\hline
\left(  \overline{1},\tilde{0}\right)  & \left(  \overline{1},\tilde{0}\right)
& \left(  \overline{0},\tilde{0}\right)  & \left(  \overline{1},\tilde
{1}\right)  & \left(  \overline{0},\tilde{1}\right) \\\hline
\left(  \overline{0},\tilde{1}\right)  & \left(  \overline{0},\tilde{1}\right)
& \left(  \overline{1},\tilde{1}\right)  & \left(  \overline{0},\tilde
{0}\right)  & \left(  \overline{1},\tilde{0}\right) \\\hline
\left(  \overline{1},\tilde{1}\right)  & \left(  \overline{1},\tilde{1}\right)
& \left(  \overline{0},\tilde{1}\right)  & \left(  \overline{1},\tilde
{0}\right)  & \left(  \overline{0},\tilde{0}\right) \\\hline
\end{array}
\]
Denoting by $\lambda_{0}=\left(  \overline{0},\tilde{0}\right)  $,
$\lambda_{1}=\left(  \overline{1},\tilde{0}\right)  $, $\lambda_{2}=\left(
\overline{0},\tilde{1}\right)  $ y $\lambda_{3}=\left(  \overline{1},\tilde
{1}\right)  $ we see that the multiplication table of the Klein group, is
given by
\begin{equation}%
\begin{array}
[c]{|c||c|c|c|c|}\hline
D_{4} & \lambda_{0} & \lambda_{1} & \lambda_{2} & \lambda_{3}\\\hline\hline
\lambda_{0} & \lambda_{0} & \lambda_{1} & \lambda_{2} & \lambda_{3}\\\hline
\lambda_{1} & \lambda_{1} & \lambda_{0} & \lambda_{3} & \lambda_{2}\\\hline
\lambda_{2} & \lambda_{2} & \lambda_{3} & \lambda_{0} & \lambda_{1}\\\hline
\lambda_{3} & \lambda_{3} & \lambda_{2} & \lambda_{1} & \lambda_{0}\\\hline
\end{array}
\label{eq:klein}%
\end{equation}

\section{\textbf{Appendix 2: Transgression forms}}

\subsection{\textbf{Calculation of }$Q^{(5)}(A_{1},\bar{A})$}

From equations (\ref{eq:conc}), we have
\[
A_{t}(A_{1},\bar{A})=\bar{A}+t(A_{1}-\bar{A})=t\omega,
\]
so that
\begin{align*}
F_{t}(A_{1},\bar{A})  &  =dA_{t}(A_{1},\bar{A})+\frac{1}{2}\left[  A_{t}%
(A_{1},\bar{A}),A_{t}(A_{1},\bar{A})\right]  ,\\
&  =td\omega+\frac{1}{2}\left[  t\omega,t\omega\right]  ,\\
&  =td\omega+\frac{t^{2}}{2}\left[  \omega,\omega\right]  ,
\end{align*}
with
\begin{align*}
d\omega &  =\left(  \frac{1}{2}\omega^{ab}J_{ab}\right)  \propto J_{ab}\\
\left[  \omega,\omega\right]  =\frac{1}{4}\omega^{ab}\omega^{cd}\left[
J_{ab},J_{cd}\right]   &  \propto J_{ab}.
\end{align*}
Since there are no invariant tensors of the form $\left\langle J_{ab}%
,J_{cd},J_{ef}\right\rangle ,$ we have
\[
Q^{(5)}\left(  A_{1},\bar{A}\right)  =3\int_{0}^{1}dt\left\langle
\omega\left(  t{d\omega}+\frac{t^{2}}{2}{\left[  \omega,\omega\right]
}\right)  ^{2}\right\rangle =0.
\]

\subsection{\textbf{Calculation of }$Q^{(5)}\left(  A_{2},A_{1}\right)  $}

Similarly to the previous case, we have
\[
A_{t}(A_{2},A_{1})=A_{1}+t(A_{2}-A_{1})=\omega+te,
\]
so that
\begin{align}
F_{t}(A_{2},A_{1})  &  =dA_{t}(A_{2},A_{1})+\frac{1}{2}\left[  A_{t}%
(A_{2},A_{1}),A_{t}(A_{2},A_{1})\right]  ,\nonumber\\
&  =d\omega+tde+\frac{1}{2}\left[  \omega+te,\omega+te\right]  ,\nonumber\\
&  =d\omega+tde+\frac{1}{2}\left[  \omega,\omega\right]  +\frac{t}{2}\left[
\omega,e\right]  +\frac{t}{2}\left[  e,\omega\right]  +\frac{t^{2}}{2}\left[
e,e\right]  , \label{c32-1}%
\end{align}
from where
\begin{align*}
R=d\omega+\frac{1}{2}\left[  \omega,\omega\right]   &  \propto J_{ab},\\
T=de+\left[  \omega,e\right]   &  \propto P_{a},\\
\left[  e,e\right]  =\frac{1}{\ell^{2}}e^{a}e^{b}\left[  P_{a},P_{b}\right]
&  \propto Z_{ab}.
\end{align*}
and therefore
\begin{align*}
Q^{(5)}\left(  A_{2},A_{1}\right)   &  =3\int_{0}^{1}dt\left\langle e\left(
R+t{T}+\frac{t^{2}}{2}\left[  e,e\right]  \right)  ^{2}\right\rangle ,\\
&  =3\int_{0}^{1}dt\left\langle e\left(  R+\frac{t^{2}}{2}\left[  e,e\right]
\right)  ^{2}\right\rangle ,\\
Q^{(5)}\left(  A_{2},A_{1}\right)   &  =3\int_{0}^{1}dt\left\langle
R^{2}e+t^{2}R\left[  e,e\right]  e+\frac{t^{4}}{4}\left[  e,e\right]
^{2}e\right\rangle
\end{align*}
so that
\begin{align}
&  Q^{(5)}\left(  A_{2},A_{1}\right) \nonumber\\
&  =3\left\langle R^{2}e+\frac{1}{3}R\left[  e,e\right]  e+\frac{1}{20}\left[
e,e\right]  ^{2}e\right\rangle ,\nonumber\\
&  =3\left(  \frac{1}{4\ell}R^{ab}R^{cd}e^{e}\left\langle J_{ab},J_{cd}%
,P_{e}\right\rangle +\frac{1}{6\ell^{3}}R^{ab}e^{c}e^{d}e^{e}\left\langle
J_{ab},\left[  P_{c},P_{d}\right]  ,P_{e}\right\rangle \right. \nonumber\\
&  +\left.  \frac{1}{20\ell^{5}}e^{a}e^{b}e^{c}e^{d}e^{e}\left\langle \left[
P_{a},P_{b}\right]  ,\left[  P_{c},P_{d}\right]  ,P_{e}\right\rangle \right)
,\nonumber\\
&  =3\left(  \frac{1}{4\ell}R^{ab}R^{cd}e^{e}\left\langle J_{ab},J_{cd}%
,P_{e}\right\rangle +\frac{1}{6\ell^{3}}R^{ab}e^{c}e^{d}e^{e}\left\langle
J_{ab},Z_{cd},P_{e}\right\rangle +\frac{1}{20\ell^{5}}e^{a}e^{b}e^{c}%
e^{d}e^{e}\left\langle Z_{ab},Z_{cd},P_{e}\right\rangle \right)  ,\nonumber\\
&  =3\,\varepsilon_{abcde}\left(  \frac{1}{4\ell}\left(  \alpha_{0}+\alpha
_{1}\right)  R^{ab}R^{cd}e^{e}+\frac{1}{6\ell^{3}}\left(  \alpha_{1}%
+\alpha_{2}\right)  R^{ab}e^{c}e^{d}e^{e}+\frac{1}{20\ell^{5}}\left(
\alpha_{2}+\alpha_{3}\right)  e^{a}e^{b}e^{c}e^{d}e^{e}\right)  .
\label{eq:a2a1}%
\end{align}

If we demand that the constants $\alpha_{1},\alpha_{2}$ and $\alpha_{3}$
satisfy the conditions
\begin{equation}%
\begin{array}
[c]{l}%
\alpha_{0}+\alpha_{1}=0,\\
\alpha_{1}-\alpha_{2}=0,\\
\alpha_{2}-\alpha_{3}=0,
\end{array}
\end{equation}
we find%
\begin{equation}
\left(  \alpha_{0},\alpha_{1},\alpha_{2},\alpha_{3}\right)  =\alpha_{0}\left(
1,-1,-1,-1\right)  . \label{a1}%
\end{equation}
In this case the invariant tensors (\ref{eq:tensinvcesc}) are given by
\begin{equation}%
\begin{array}
[c]{l}%
\left\langle J_{ab},J_{cd},Z_{e}\right\rangle =\alpha_{0}\,\varepsilon
_{abcde},\\
\left\langle J_{ab},Z_{cd},P_{e}\right\rangle =-\alpha_{0}\,\varepsilon
_{abcde},\\
\left\langle Z_{ab},Z_{cd},P_{e}\right\rangle =-\alpha_{0}\,\varepsilon
_{abcde}.
\end{array}
\label{c6}%
\end{equation}
and $Q^{(5)}\left(  A_{2},A_{1}\right)  $ takes the form
\begin{equation}
Q^{(5)}\left(  A_{2},A_{1}\right)  =3\,\varepsilon_{abcde}\left(  -\frac
{1}{3\ell^{3}}\alpha_{0}\,R^{ab}e^{c}e^{d}e^{e}-\frac{1}{10\ell^{5}}\alpha
_{0}\,e^{a}e^{b}e^{c}e^{d}e^{e}\right)  .
\end{equation}

Note that the form of transgression (\ref{eq:a2a1}) contains the same terms as
the Lagrangian of Lovelock on 5 dimensions for any value of the coefficients
$\alpha.$ \ We must not confuse the coefficients $\alpha$ with the
coefficients of Lovelock. If we denote by $\beta$ the coefficients of
Lovelock, then the relationship between the expansion coefficients $\alpha$
and the coefficients of Lovelock, is given by
\begin{equation}%
\begin{array}
[c]{l}%
\beta_{0}=\left(  \alpha_{0}+\alpha_{1}\right)  /2,\\
\beta_{1}=\left(  \alpha_{1}+\alpha_{2}\right)  /(3\ell^{2}),\\
\beta_{2}=\left(  \alpha_{2}+\alpha_{3}\right)  /(10\ell^{4}).
\end{array}
\end{equation}

\subsection{\textbf{Calculation of} $Q^{(5)}\left(  A,A_{2}\right)  $}

In this case we have
\[
A_{t}(A,A_{2})=A_{2}+t(A-A_{2})=\omega+e+t(k+h),
\]
so that
\begin{align*}
&  F_{t}(A,A_{2})\\
&  =d\omega+de+t(dk+dh)+\frac{1}{2}\left[  \omega,\omega\right]  +\left[
\omega,e\right]  +t\left[  \omega,k\right]  +t\left[  \omega,h\right] \\
&  +\frac{1}{2}\left[  e,e\right]  +t\left[  e,h\right]  +t\left[  k,e\right]
+\frac{t^{2}}{2}\left[  k,k\right]  +t^{2}\left[  k,h\right]  +\frac{t^{2}}%
{2}\left[  h,h\right]  ,\\
&  =d\omega+\frac{1}{2}\left[  \omega,\omega\right]  +de+\left[
\omega,e\right]  +\frac{1}{2}\left[  e,e\right]  +t\left(  dk+\left[
\omega,k\right]  +dh+\left[  \omega,h\right]  +\left[  k,e\right]  +\left[
e,h\right]  \right) \\
&  +t^{2}\left(  \frac{1}{2}\left[  k,k\right]  +\frac{1}{2}\left[
h,h\right]  +\left[  k,h\right]  \right)  ,\\
&  =R+T+\frac{1}{2}\left[  e,e\right]  +t\left(  \mathrm{D}_{\omega
}k+\mathrm{D}_{\omega}h+\left[  k,e\right]  +\left[  e,h\right]  \right)
+\frac{t^{2}}{2}\left(  \left[  k,k\right]  +\left[  h,h\right]  +2\left[
k,h\right]  \right)  ,
\end{align*}
from where
\[%
\begin{array}
[c]{ll}%
R=d\omega+\frac{1}{2}\left[  \omega,\omega\right]  \propto J_{ab}, & \left[
k,e\right]  \propto Z_{a},\\
T=de+\left[  \omega,e\right]  \propto P_{a}, & \left[  e,h\right]  \propto
J_{ab},\\
\left[  e,e\right]  \propto Z_{ab}, & \left[  k,k\right]  \propto J_{ab},\\
\mathrm{D}_{\omega}k=dk+\left[  \omega,k\right]  \propto Z_{ab}, & \left[
h,h\right]  \propto Z_{ab},\\
\mathrm{D}_{\omega}h=dh+\left[  \omega,h\right]  \propto Z_{a}, & \left[
k,h\right]  \propto P_{a},
\end{array}
\]
and therefore
\begin{align*}
&  Q^{(5)}\left(  A,A_{2}\right) \\
&  =3\,\alpha_{0}\,\varepsilon_{abcde}\left(  \frac{1}{2\ell}k^{ab}R^{cd}%
T^{e}\right.  +\frac{1}{6\ell}R^{ab}k^{cd}k_{f}^{e}h^{f}+\frac{1}{2\ell^{3}%
}k^{ab}T^{c}e^{d}e^{e}+\frac{1}{4\ell}k^{ab}T^{c}\mathrm{D}_{\omega}k^{de}\\
&  +\frac{1}{2\ell^{3}}k^{ab}T^{c}e^{d}h^{e}-\frac{1}{6\ell}k^{ab}T^{c}%
k_{f}^{d}k^{fe}-\frac{1}{6\ell^{3}}k^{ab}T^{c}h^{d}h^{e}+\frac{1}{6\ell^{3}%
}k^{ab}e^{c}e^{d}k_{f}^{e}h^{f}\\
&  +\frac{1}{8\ell}k^{ab}\mathrm{D}_{\omega}k^{cd}k_{f}^{e}h^{f}+\frac
{1}{4\ell^{3}}k^{ab}e^{c}h^{d}k_{f}^{e}h^{f}-\frac{1}{10\ell}k^{ab}k_{f}%
^{c}k^{fd}k_{g}^{e}h^{g}-\frac{1}{10\ell^{3}}k^{ab}h^{c}h^{d}k_{f}^{e}k^{f}\\
&  -\frac{1}{4\ell}R^{ab}R^{cd}h^{e}-\frac{1}{3\ell^{5}}e^{a}h^{b}e^{c}%
h^{d}h^{e}-\frac{1}{20\ell}k_{f}^{a}k^{fb}k_{g}^{c}k^{gd}h^{e}-\frac{1}%
{2\ell^{3}}R^{ab}e^{c}h^{d}h^{e}\\
&  +\left.  \frac{1}{4\ell^{3}}k_{f}^{a}k^{fb}e^{c}h^{d}h^{e}+\frac{1}{6\ell
}R^{ab}k_{f}^{c}k^{fd}h^{e}\right)  .
\end{align*}

\end{document}